%% file: 0_main_arxiv.tex
\documentclass[nonacm,acmtog,screen]{acmart}

\author{Caoliwen Wang}
\authornote{Equal contribution}
\email{wclw1021@gmail.com}
\orcid{0009-0004-3822-9094}
\affiliation{%
  \institution{University of British Columbia}
  \country{Canada}
}

\author{Minghao Guo}
\authornotemark[1]
\authornote{Corresponding author}
\email{guomh2014@gmail.com}
\orcid{0000-0003-3408-4997}
\affiliation{%
  \institution{MIT CSAIL}
  \country{USA}
}

\author{Siyuan Chen}
\authornotemark[1]
\email{chensiyuan030105@gmail.com}
\orcid{0009-0009-4309-862X}
\affiliation{%
  \institution{University of British Columbia}
  \country{Canada}
}

\author{Heng Zhang}
\email{mediosrity@gmail.com}
\orcid{0000-0002-2836-2965}
\affiliation{%
  \institution{University of British Columbia}
  \country{Canada}
}

\author{Mengdi Wang}
\email{m.wang.13@outlook.com}
\orcid{0000-0001-5757-3510}
\affiliation{%
  \institution{Georgia Institute of Technology}
  \country{USA}
}

\author{Xingyu Ni}
\email{xingyu.ni@inria.fr}
\orcid{0000-0003-1127-2848}
\affiliation{%
  \institution{Inria}
  \country{France}
}

\author{Hanson Sun}
\email{jsun33@student.ubc.ca}
\orcid{0009-0002-6228-905X}
\affiliation{%
  \institution{University of British Columbia}
  \country{Canada}
}

\author{Kunyi Wang}
\email{kunyiwang252750@gmail.com}
\orcid{0009-0007-2832-0574}
\affiliation{%
  \institution{University of British Columbia}
  \country{Canada}
}

\author{Zherong Pan}
\email{zherong.pan.usa@gmail.com}
\orcid{0000-0001-9348-526X}
\affiliation{%
  \institution{Meta}
  \country{USA}
}

\author{Kui Wu}
\email{walker.kui.wu@gmail.com} 
\orcid{0000-0003-3326-7943}
\affiliation{%
  \institution{Independent Researcher}
  \country{USA}
}

\author{Lingjie Liu}
\email{lingjie.liu@seas.upenn.edu}
\orcid{0000-0003-4301-1474}
\affiliation{%
  \institution{University of Pennsylvania}
  \country{USA}
}

\author{Yin Yang}
\email{yangzzzy@gmail.com}
\orcid{0000-0001-7645-5931}
\affiliation{%
  \institution{University of Utah}
  \country{USA}
}

\author{Chenfanfu Jiang}
\email{chenfanfu.jiang@gmail.com}
\orcid{0000-0003-3506-0583}
\affiliation{%
  \institution{University of California Los Angeles}
  \country{USA}
}

\author{Taku Komura}
\email{taku@cs.hku.hk}
\orcid{0000-0002-2729-5860}
\affiliation{%
  \institution{University of Hong Kong}
  \country{Hong Kong}
}

\author{Wojciech Matusik}
\email{wojciech@csail.mit.edu}
\orcid{0000-0003-0212-5643}
\affiliation{%
  \institution{MIT CSAIL}
  \country{USA}
}

\author{Peter Yichen Chen}
\authornotemark[2]
\email{pyc@csail.mit.edu}
\orcid{0000-0003-1919-5437}
\affiliation{%
  \institution{University of British Columbia}
  \country{Canada}
}

\usepackage[ruled]{algorithm2e} % For algorithms

\SetAlFnt{\small}
\SetAlCapFnt{\small}
\SetAlCapNameFnt{\small}
\SetAlCapHSkip{0pt}

\usepackage{enumitem}
\usepackage{makecell}
\usepackage{wrapfig}
\usepackage{subcaption}
\usepackage{enumitem}
\usepackage{makecell}
\usepackage{wrapfig}
\usepackage{subcaption}
\usepackage{setspace}
\usepackage{bm}
\usepackage{algorithmic}

\definecolor{wmd_navyblue}{HTML}{000080}

\newcommand{\methodname}{WorldParticle}

\usepackage{tabularx}
\usepackage{multirow}

\begin{document}

%%
%% The "title" command has an optional parameter,
%% allowing the author to define a "short title" to be used in page headers.
% \title{our model: Transformer-based Simulation of Lagrangian Dynamics}
% \title{our model: Unified Simulation of Lagrangian Dynamics via Transformers}
\title{\methodname: Unified World Simulation of Lagrangian
Particle Dynamics via Transformer}
\acmSubmissionID{1695}

%%
%% The "author" command and its associated commands are used to define
%% the authors and their affiliations.
%% Of note is the shared affiliation of the first two authors, and the
%% "authornote" and "authornotemark" commands
%% used to denote shared contribution to the research.
% \author{Ben Trovato}
% \authornote{Both authors contributed equally to this research.}
% \email{trovato@corporation.com}
% \orcid{1234-5678-9012}
% \author{G.K.M. Tobin}
% \authornotemark[1]
% \email{webmaster@marysville-ohio.com}
% \affiliation{%
%   \institution{Institute for Clarity in Documentation}
%   \city{Dublin}
%   \state{Ohio}
%   \country{USA}
% }

% \author{Lars Th{\o}rv{\"a}ld}
% \affiliation{%
%   \institution{The Th{\o}rv{\"a}ld Group}
%   \city{Hekla}
%   \country{Iceland}}
% \email{larst@affiliation.org}
%%
%% By default, the full list of authors will be used in the page
%% headers. Often, this list is too long, and will overlap
%% other information printed in the page headers. This command allows
%% the author to define a more concise list
%% of authors' names for this purpose.
\renewcommand{\shortauthors}{Wang \& Guo \& Chen et al.}

%%
%% The abstract is a short summary of the work to be presented in the
%% article.
\begin{abstract}
A unified simulator that can model diverse physical phenomena without solver-specific redesign is a long-standing goal across simulation science. We present a learning-based particle simulator built on a single transformer architecture to model cloth, elastic solds, Newtonian and non-Newtonian fluids, granular materials, and molecular dynamics. Our model follows a prediction-correction design on a shared Lagrangian particle representation. An explicit predictor first advances particles under the known external forces, producing an intermediate state that captures externally driven motion but not inter-particle interactions. A learned corrector then predicts the residual position and velocity updates through three stages: a particle tokenizer that encodes local particle-particle, particle-boundary, and topology-guided interactions; a super-token encoder that hierarchically merges particle tokens into a compact set of super tokens via alternating self-attention and token merging; and a super-token decoder that lifts these super tokens back to particle resolution through cross-attention to predict per-particle position and velocity corrections. Progressive token merging reduces the attention cost at successive encoder layers by halving the token count at each level, and the decoder communicates through the compact super-token set rather than full particle-to-particle attention. Across the six dynamics categories, the same architecture generalizes to unseen materials, boundary configurations, initial conditions, and external forces. We further demonstrate downstream interactive control, inverse design, and learning from real-world manipulation data, reducing the need for per-phenomenon solver engineering.
\end{abstract}

%%
%% The code below is generated by the tool at http://dl.acm.org/ccs.cfm.
%% Please copy and paste the code instead of the example below.
%%

% \ccsdesc[300]{Computing methodologies~Rendering}
% \ccsdesc[300]{Computing methodologies~Machine learning approaches}

%%
%% Keywords. The author(s) should pick words that accurately describe
%% the work being presented. Separate the keywords with commas.

% \received{20 February 2007}
% \received[revised]{12 March 2009}
% \received[accepted]{5 June 2009}
\begin{teaserfigure}
 \centering
 \includegraphics[width=0.95\textwidth]{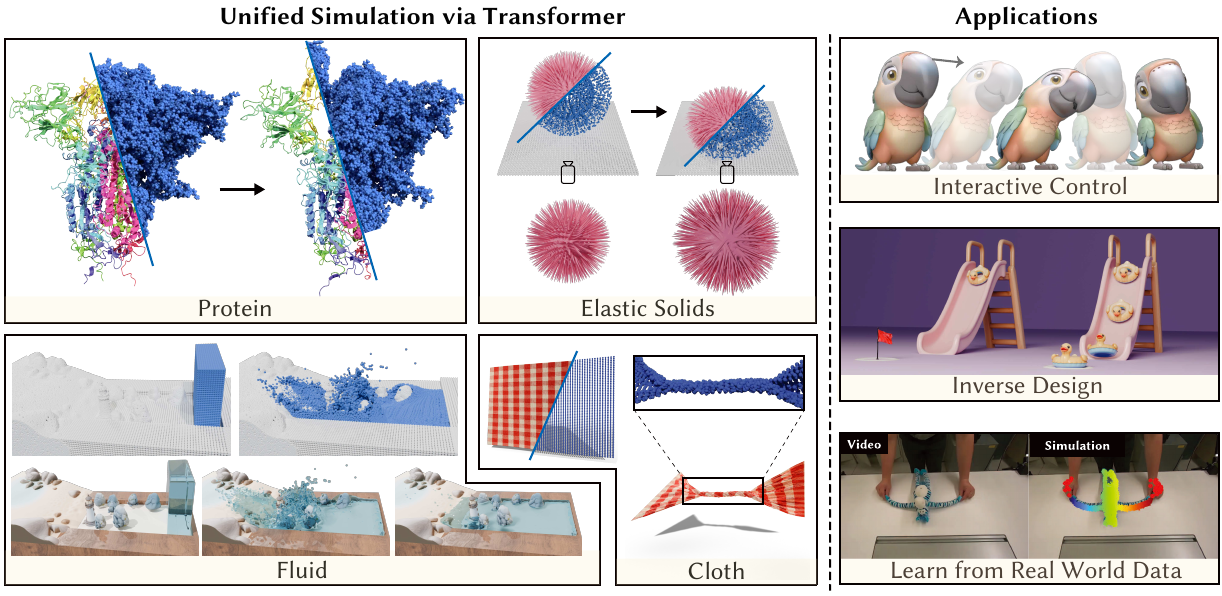}
 \vspace*{-8pt}
 \caption{\textbf{We propose a unified transformer-based neural simulator for Lagrangian particle dynamics.} 
 \textit{Left:} Our model handles diverse particle systems, including proteins, elastic solids, fluids, and cloth, within a single architecture. \textit{Right:} The learned simulator supports downstream tasks including interactive control, inverse design, and learning from real-world observations.}
 % \textit{Left:} Our model simulate complex Lagrangian particle dynamics across a wide range of systems, including proteins, deformables, fluids, and cloth, capturing intricate interactions and realistic behaviors. \textit{Right:} We demonstrate a variety of downstream applications, such as interactive control, inverse design, and learning from real-world manipulation videos. 
 % \mh{Zoom in a bit the two bottom-view figures for the hair ball example}
 % \lw{needs to change deformable object to deformables}
 % }
 \label{fig:teaser}
\end{teaserfigure}
%%
%% This command processes the author and affiliation and title
%% information and builds the first part of the formatted document.
\maketitle

\input{1_introduction}
\input{2_related_work_v2}
\input{3_method_v2}
% \input{4_experiments}
\input{4_experiments_v2}
\input{5_ablation}
\input{6_application}
\input{7_conclusion}
\bibliographystyle{ACM-Reference-Format}
\bibliography{bibfile}

\clearpage
\input{extra_figs_v2}
% \newpage
\appendix
\input{Appendix_arxiv}

\end{document}

%% file: 1_introduction.tex
\section{Introduction}

Building a \emph{unified simulation}, a single framework that captures the full diversity of physical phenomena, has long been pursued across simulation science, from computer graphics and engineering simulation to molecular modeling, but remains elusive in practice. Different phenomena are traditionally modeled
with different discretizations, from Eulerian grids for fluids, to
Lagrangian meshes and the finite-element method for elastic solids,
to particle samples for free-surface flow, and by different governing
equations, from Navier--Stokes to nonlinear constitutive laws,
granular rheology, and atomistic potentials, many exhibiting
strong nonlinearity and multiscale coupling. Modern engines such as
NVIDIA Newton~\cite{newton2025} push toward broader coverage by
combining several solver backends in one system, yet each phenomenon
still relies on its own dedicated solver internally. This
\emph{per-phenomenon, per-solver} paradigm makes cross-phenomenon
coupling difficult and constrains how broadly a single classical
pipeline can be applied across the diverse physics of real-world
scenes.

% ----- Paragraph 2: particle-based representation as the first axis -----
A promising direction is to unify the underlying discrete
representation. Position-based dynamics~\cite{muller2007position} uses
particles with energy constraints to simulate cloth, soft bodies, and
fluids within a single framework, at the cost of physical fidelity.
The material point method~\cite{sulsky1994particle, stomakhin2013material} adopts a hybrid
particle-grid representation and
captures fluid-solid coupling with higher accuracy, but still relies
on per-scene parameter tuning and is computationally expensive.
Despite these tradeoffs, both demonstrate that a particle-based
Lagrangian representation is a natural common language across physical
phenomena: material points, mesh vertices, SPH samples, and atoms can
all be abstracted as particles carrying position, velocity, and
per-particle physical properties. Particles, however, only unify the
representation layer; the complex nonlinearity and problem-specific
parameterization of the underlying governing equations remain, and no
analytical model covers all phenomena at once.

% ----- Paragraph 3: ML as the second axis, with a need for long-range ---
Machine learning offers a complementary answer on the modeling side:
rather than hand-designing constitutive and interaction laws, a network
can learn dynamics directly from trajectory data,
bypassing explicit
PDEs and per-scene tuning. 
Existing neural
simulators~\cite{dpinet,deepLagrange,gns,meshgraphnets}, however,
are typically designed and trained for a specific phenomenon or
equation class, and reduced-order
methods~\cite{barbic2005real, treuille2006model} further tie their learned
bases to a fixed geometry or shape family. Changing the physics often requires changing the model design.

% reduced model need to change basis for each equation, unified simulation

% ----- Paragraph 4: our model + how it addresses local-and-global ------
We propose a particle-based prediction-correction transformer whose architecture remains \emph{entirely fixed} across phenomena, only the training data changes. The same architecture simulates cloth, elastic solids, Newtonian and non-Newtonian fluids, granular materials, and molecular dynamics. Following the prediction-correction scheme common in particle simulation~\cite{muller2007position}, an explicit predictor integrates known external forces, and a learned corrector then predicts interaction corrections for position and velocity. The corrector has three stages: a \emph{particle tokenizer} encodes local interactions through learnable kernels over spatial, boundary, and topological neighborhoods; inspired by multigrid hierarchical coarsening~\cite{vanvek1996algebraic}, a \emph{super-token encoder} compresses these tokens into a compact set of super tokens by alternating self-attention and token merging, progressively halving the token count; and a \emph{super-token decoder} lifts super-token information back to full particle resolution through cross-attention, producing residual position and velocity corrections. Because both the architecture and the prediction-correction interface are equation-agnostic, the model treats each phenomenon as a different training distribution over the same input-output format, rather than requiring a new model design for each governing equation. The pipeline is trained
with an autoregressive rollout loss.

% ----- Paragraph 5: applications + sim-to-real (single tight paragraph) --
The same architecture, trained on the six Lagrangian dynamics
categories above, generalizes to unseen materials, boundary
configurations, initial conditions, and external forces.
Our model delivers
higher fidelity and more stable long-horizon rollouts than existing neural simulators while covering a broader range of physical phenomena in a single architecture. 
Beyond
forward simulation, 
we demonstrate three downstream applications: interactive control of deformable objects under user-specified
forces; inverse design of friction parameters via
differentiable rollout;
and
real-world manipulation prediction on particle trajectories extracted
with PhysTwin~\cite{jiang2025phystwin} using 3D Gaussian
Splatting~\cite{kerbl20233d}.
Together, our model handles phenomena whose classical counterparts each require dedicated FEM, SPH, or MPM solvers, using a single differentiable architecture and workflow, at the cost of some per-phenomenon accuracy compared to those specialized methods.
Our contributions are:
\begin{itemize}[leftmargin=*]
    \item A particle-based prediction-correction transformer that simulates cloth, elastic solids, Newtonian and
    non-Newtonian fluids, granular materials, and molecular dynamics, in a single architecture.
    \item A particle tokenizer for local neighborhood interactions combined with a super-token encoder-decoder for global communication through progressive token merging.
    \item Experiments across six dynamics categories showing lower rollout errors and more stable long-horizon predictions than existing neural simulators, together with three applications: interactive control, inverse design, and real-world manipulation prediction.
\end{itemize}

%% file: 2_related_work_v2.tex
\section{Related work}
We review three lines of work: learning-based physics simulators that operate on particles and meshes, reduced-order and operator-learning methods that accelerate simulation through compact representations, and transformer architectures applied to graphics.
%~\zrpan{Shall we talk about conventional simulation techniques with local-global communication, e.g. ADMM, multigrid?}\md{I think multigrid is only related philosophically, so no need for citation here.}~\zrpan{Fine with me. Comment deleted.}

\paragraph{Learning-based physics simulation.}
Learning-based simulators have progressed from object-level prediction to models operating directly on particles and meshes. Interaction networks~\cite{interactionnetworks} and neural physics engines~\cite{npe} predict dynamics of predefined object sets but require manually specified interaction graphs. SPNets~\cite{spnets} and DPI-Net~\cite{dpinet} learn particle-level updates across multiple material types. Continuous-convolution networks~\cite{deepLagrange} replace hand-crafted SPH kernels with learned ones. GNS~\cite{gns} and MeshGraphNets~\cite{meshgraphnets} learn local message passing on particle radius graphs and simulation meshes, respectively. Domain-specific neural simulators further exploit physical structure for Eulerian-fluid acceleration~\cite{accelEulerian}, smoke synthesis~\cite{smokecnn}, fluid super-resolution~\cite{tempogan}, cloth deformation~\cite{neuralcloth,pbns}, garment collision handling~\cite{senc}, and constitutive-law learning within differentiable solvers~\cite{nclaw}. A common limitation is that global information must propagate through many local hops. Our work retains the Lagrangian particle representation and local neighborhood structure, but introduces a super-token encoder-decoder that provides global communication in every correction step without deep message-passing chains.

\paragraph{Reduced-order modeling and neural operators.}
A complementary direction accelerates simulation by compressing the state or the solution operator. Latent-space models encode high-dimensional fields into compact coordinates and learn their temporal evolution~\cite{deepfluids,latentspacephysics,deepflowcontrol}. Neural-field and subspace methods learn reduced representations spanning specific geometry families or solution classes~\cite{crom,licrom,shapespacespectra,precisegradients,fastsubspacefluid,lowrankkoopman,factorizeddmd,neuralstressfields,simplicits,odddc}, but new geometries or physics typically require rebuilding the basis. Neural operators learn maps between function spaces~\cite{pinns,deeponet,fno}, with extensions to multiscale problems~\cite{mgno,mgtfno,ffno,cno}, complex geometries~\cite{gino,beno}, and physical consistency~\cite{pino}; they predict fields over PDE families rather than evolving individual particle states. Our model performs particle-level evolution: super tokens serve as intermediate communication variables regenerated from the full dynamical state at every timestep, rather than a fixed reduced basis or a learned solution operator.

\paragraph{Transformers in graphics and simulation.}
Transformers provide data-dependent, nonlocal communication on geometric domains. In point-cloud processing, self-attention supports geometric reasoning, shape completion, and pretraining~\cite{pointtransformer,pointr,pointbert,pointmae,pointtransformerv2,octformer,swin3d,pointtransformerv3}. In rendering, attention aggregates multiview and ray information for view synthesis and global illumination~\cite{ibrnet,viewformer,lvsm,renderformer}. Transformers have also been applied to physics surrogates for capturing long-range dependencies on general domains~\cite{transolver,p3d}. Video and generative world models produce visually plausible motion but operate on pixels or latent frames rather than explicit particle states~\cite{physctrl}. Our method applies transformer attention to particle-level simulation, covering a wide range of physical phenomena within a single architecture.

%% file: 3_method_v2.tex
\section{Method}
\label{sec:method}
We model one step of Lagrangian particle dynamics as a \emph{prediction-correction} procedure~\cite{muller2007position, ladicky2015data}.
An explicit prediction step integrates known external forces, producing an intermediate state that accounts for external acceleration but not for inter-particle interactions.
A learned correction step then predicts the interaction correction.
The corrector consists of three components: a \emph{particle tokenizer} that encodes particle-particle, particle-boundary, and topological interactions, a \emph{super-token encoder} that compresses particle tokens into dynamically generated super tokens, and a \emph{super-token decoder} that lifts super-token information back to particle resolution to produce position and velocity corrections.
Together, they separate communication into two scales: the tokenizer captures \emph{local} interactions, while the encoder and decoder handle \emph{global} coupling through the super tokens.

\begin{figure*}[t]
    \centering
    \includegraphics[width=0.98\textwidth]{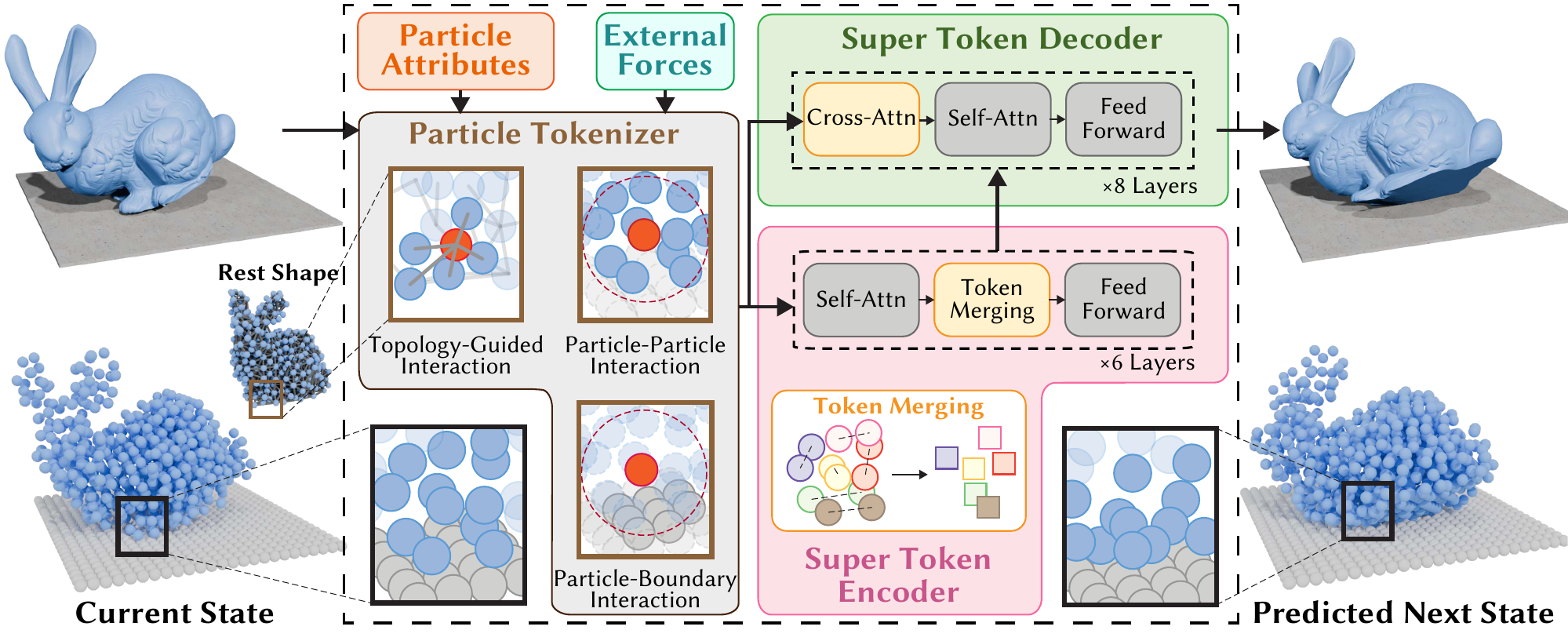}
    \vspace{-1em}
    \caption{\textbf{Prediction-correction particle transformer.} Given the intermediate state from the prediction step (Eq.~\ref{eq:predictor}), the correction step proceeds in three stages within one timestep: the \emph{particle tokenizer} encodes local interactions into per-particle tokens, the \emph{super-token encoder} compresses them into dynamically generated super tokens via self-attention and token merging, and the \emph{super-token decoder} refines particle tokens through alternating self-attention and cross-attention to produce position and velocity corrections (Eq.~\ref{eq:corrector}).}
    \label{fig:pipeline}
\end{figure*}

Let $\bm{X}_t,\bm{V}_t,\bm{F}_t\in\mathbb{R}^{N\times3}$ denote positions, velocities, and external forces of the $N$ simulated particles, and let $\bm{C}\in\mathbb{R}^{N\times C_p}$ denote their per-particle attributes such as mass, material parameters, or atom types.
Separately, $N_b$ static boundary particles $(\bm{X}^{\textsc{b}},\bm{C}^{\textsc{b}})$ represent walls, obstacles, or contact surfaces, with attributes such as normals and friction coefficients.
$\mathcal{T}$ denotes rest-shape topology when available (e.g., for cloth and elastic solids).
The prediction step integrates the external forces over one timestep,
\begin{align}
\tilde{\bm{V}}_t
&= \bm{V}_t+\Delta t\,\bm{M}^{-1}\bm{F}_t,
&
\tilde{\bm{X}}_t
&= \bm{X}_t+\tfrac{\Delta t}{2}(\bm{V}_t+\tilde{\bm{V}}_t),
\label{eq:predictor}
\end{align}
where $\bm{M}$ is the diagonal mass matrix assembled from $\bm{C}$.
The correction step takes the predicted state and outputs residual updates:
\begin{align}
(\Delta\bm{X}_t,\Delta\bm{V}_t)
=\mathcal{F}_\theta(\tilde{\bm{X}}_t,\tilde{\bm{V}}_t,\bm{C},\mathcal{T},\bm{X}^{\textsc{b}},\bm{C}^{\textsc{b}}),
\label{eq:corrector}
\end{align}
yielding $\bm{X}_{t+\Delta t}=\tilde{\bm{X}}_t+\Delta\bm{X}_t$ and $\bm{V}_{t+\Delta t}=\tilde{\bm{V}}_t+\Delta\bm{V}_t$. Predicting both residuals lets the model represent effects from implicit solves, contact projections, pressure coupling, and damping that cannot be expressed as a single explicit acceleration. Inputs unavailable in a domain, such as $\mathcal{T}$ for fluids or boundary particles for proteins, are zero-filled so that the architecture remains fixed.
% \lw{Actually, if we do not have Topology, maybe the embedding be dim $\times$ 2 not for dim $\times$3}.
Throughout, bold uppercase letters denote ordered sets of particles (e.g., $\bm{X}$) and bold lowercase with subscript~$i$ denotes the $i$-th element (e.g., $\bm{x}_i$).
A superscript $^{(\ell)}$ indicates the encoder or decoder layer; when omitted it defaults to layer~0 (the original particle resolution), so in particular $N^{(0)}\!=\!N$.
The following subsections describe one timestep and drop the time subscript; lowercase symbols such as $(\tilde{\bm{x}}_i,\tilde{\bm{v}}_i,\bm{c}_i)$ denote the $i$-th row of $(\tilde{\bm{X}}_t,\tilde{\bm{V}}_t,\bm{C})$.
The components of $\mathcal{F}_\theta$ are shown in Fig.~\ref{fig:pipeline}.

% ------------------------------------------------------------
\subsection{Particle Tokenizer}
\label{sec:tokenizer}

The particle tokenizer encodes local interactions before global attention. It follows the locality principle underlying particle methods: interactions are evaluated over compact neighborhoods, while the filters inside those neighborhoods are learned from data rather than prescribed as SPH kernels~\cite{koschier2022survey}. For particle $i$, we use up to three supports: a spatial neighborhood $\mathcal{N}^{\textsc{s}}_i$ from radius search around $\tilde{\bm{x}}_i$, a boundary neighborhood $\mathcal{N}^{\textsc{b}}_i$ from nearby boundary samples, and a topological neighborhood $\mathcal{N}^{\textsc{t}}_i$ defined by the adjacency of the rest-shape tessellation $\mathcal{T}$.
Each branch aggregates neighbor contributions through a learnable kernel $\bm{W}_k$:
\begin{align}
\bm{a}^{k}_{i}
&=
\sum_{j\in\mathcal{N}^{k}_{i}}
\bm{W}_k\!\bigl(\bm{r}^{k}_{ij}\bigr)\,
\bm{u}^{k}_{ij},
\qquad k\in\{\textsc{s},\textsc{b},\textsc{t}\},
\label{eq:cconv}
\end{align}
where $\bm{W}_k$ maps a relative displacement to a mixing matrix, parameterized as a learnable 3D grid with trilinear interpolation and compact support of radius~\cite{deepLagrange} (details in the appendix). Each branch defines displacement $\bm{r}^k_{ij}$ and feature $\bm{u}^k_{ij}$ as:
\begin{equation}
\label{eq:branch_defs}
\begin{aligned}
\textit{Particle-Particle:}\quad
&\bm{r}^{\textsc{s}}_{ij}=\tilde{\bm{x}}_{j}-\tilde{\bm{x}}_{i},
&\bm{u}^{\textsc{s}}_{ij}&=[\tilde{\bm{v}}_{j},\,\bm{c}_j],\\
\textit{Particle-Boundary:}\quad
&\bm{r}^{\textsc{b}}_{ij}=\bm{x}^{\textsc{b}}_{j}-\tilde{\bm{x}}_{i},
&\bm{u}^{\textsc{b}}_{ij}&=\bm{c}^{\textsc{b}}_{j},\\
\textit{Topology-Guided:}\quad
&\bm{r}^{\textsc{t}}_{ij}={\bm{x}}^0_{j}-{\bm{x}}^0_{i},
&\bm{u}^{\textsc{t}}_{ij}&=[\tilde{\bm{v}}_{j},\,\bm{c}_j,\,\bm{x}^0_j\!-\!\bm{x}^0_i],
\end{aligned}
\end{equation}
where $\bm{x}^0_i$ denotes the particle's rest-pose position.
% \lw{Here, 1 is the same as Deep Lagrange Fluid}
% Thus the spatial branch captures current proximity, the boundary branch exposes nearby surface geometry, and the topological branch uses rest-space support while passing current deformation as a feature.
The particle token $\bm{h}_i$ concatenates all available branches with the particle's own state:
\begin{align}
\bm{h}_i &= \mathrm{MLP}\bigl([\bm{a}^{\textsc{s}}_i,\,\bm{a}^{\textsc{t}}_i,\,\bm{a}^{\textsc{b}}_i,\,\tilde{\bm{v}}_{i},\,\bm{c}_i]\bigr).
\label{eq:tokenizer}
\end{align}
% \lw{Might not like this, actually $h_i=[\bm{a}^{\textsc{s}}_i,\,\bm{a}^{\textsc{t}}_i,\,\bm{a}^{\textsc{b}}_i,MLP(1,\tilde{\bm{v}}_{i},\,\bm{c}_i)]$, and also $\bm{x}^0$ not defined}
% ~\zrpan{we still need to make subscript clear. For example, you have $\tilde{x}_t$ means particle at t-th time instance. And you have $\tilde{x}_i$ means i-th particle. Need clarify, which is easy to fix.}

% ------------------------------------------------------------
\subsection{Super-Token Encoder}
\label{sec:encoder}

\begin{figure}
    \centering
    \includegraphics[width=0.48\textwidth]{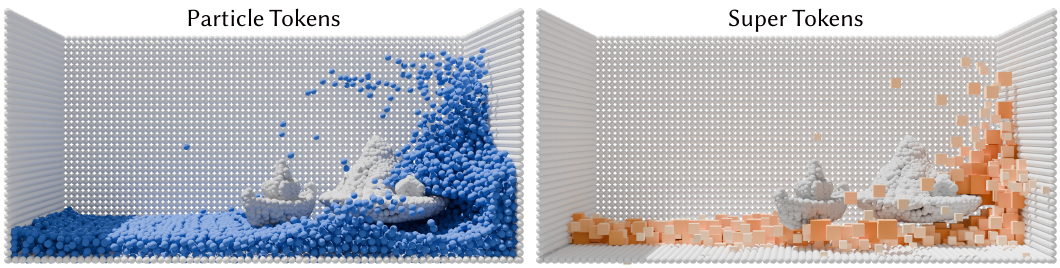}
    \vspace*{-2em}
    \caption{\textbf{Token merging visualization.} Particle tokens (blue spheres, left) are merged into super tokens (yellow cubes, right). Cube size and color intensity reflect how many original particles each super token represents.}
    \label{fig:super token}
\end{figure}

The particle tokens $\{\bm{h}_i\}$ capture local interactions, but many physical phenomena, such as pressure waves, elastic stresses, and molecular motion, require global, long-range communication.
Self-attention provides such coupling~\cite{attentionisallyouneed},
but applying it repeatedly over all $N$ particles is computationally expensive.
Inspired by the hierarchical coarsening principle of multi-grid methods~\cite{vanvek1996algebraic},
the super-token encoder alternates global self-attention at the current resolution with \emph{token merging}~\cite{bolya2022token}, so later attention layers operate on progressively fewer tokens.

Let $\bm{H}_{\mathrm e}$ and $\tilde{\bm{X}}$ be the particle tokens and positions from the tokenizer.
Each encoder layer $\ell=1,\ldots,L_e$ first applies multi-head self-attention with 3D RoPE~\cite{rope}
to contextualize all tokens at the current resolution:
\begin{align}
\widehat{\bm{H}}_\mathrm{e}^{(\ell)} &= \text{Self-Attn}\!\bigl(\bm{H}_\mathrm{e}^{(\ell-1)},\, \tilde{\bm{X}}^{(\ell-1)}\bigr).
\label{eq:sa}
\end{align}
with $\widehat{\bm{h}}^{(\ell)}_i$ denotes the $i$-th token in $\widehat{\bm{H}}^{(\ell)}_{\mathrm e}$.
We then perform token merging, which halves the sequence from $N^{(\ell-1)}$ to $N^{(\ell)}\!=\!\lceil N^{(\ell-1)}/2 \rceil$.
Specifically, the current tokens $\{\widehat{\bm{h}}^{(\ell)}_i\}$ are split into two disjoint halves by alternating index.
Each token in the first half is greedily matched to its most cosine-similar token in the second half, and matched pairs are merged while unmatched tokens are kept as-is.
% \lw{Actually, here all the particles in the first set should be merged to the second half, some particles in the second half might still not be merged and keep as-is.}
Let $\mathcal{M}^{(\ell)}_i$ denote the set of tokens collapsed into the $i$-th output token.
The merged token and its positional anchor are computed by weighted averaging:
\begin{small}
\begin{align}
\bigl(\bm{h}^{(\ell)}_i,\, \tilde{\bm{x}}^{(\ell)}_i\bigr) = \sum_{j\in\mathcal{M}^{(\ell)}_i} \frac{m^{(\ell-1)}_j}{m^{(\ell)}_i}\, \bigl(\widehat{\bm{h}}^{(\ell)}_j,\, \tilde{\bm{x}}^{(\ell-1)}_j\bigr), \ \
\label{eq:merge}
\bm{h}^{(\ell)}_i \leftarrow \mathrm{MLP}\bigl(\bm{h}^{(\ell)}_i\bigr). 
\end{align}
\end{small}
where $m^{(\ell)}_i = \sum_{j\in\mathcal{M}^{(\ell)}_i} m^{(\ell-1)}_j$ is the multiplicity of token $i$ at layer $\ell$, initialized as $m^{(0)}_i{=}1$.
The merged tokens and positions form $\bm{H}_\mathrm{e}^{(\ell)}$ and $\tilde{\bm{X}}^{(\ell)}$ with $N^{(\ell)}$ tokens, which become the input to layer $\ell\!+\!1$.
Applying self-attention \emph{before} merging is important: similarity is computed after each token has seen current-scale context, rather than from raw local features alone.
Thus, particles with similar local neighborhoods but different global roles (e.g., one near a free surface, one deep in the interior) are less likely to be merged.

After $L_e$ layers, the remaining tokens are the \emph{super tokens}
$\widebar{\bm{H}} \triangleq \bm{H}_\mathrm{e}^{(L_e)}$
with corresponding positions
$\widebar{\bm{X}} \triangleq \tilde{\bm{X}}^{(L_e)}$, as shown in Fig.~\ref{fig:super token}. Compared to classical multigrid, which relies on local smoothing operators and achieves linear per-iteration cost, our encoder uses global self-attention at each level and retains a quadratic cost at every resolution.
The trade-off is expressiveness: global attention captures long-range coupling more directly than purely local operators, and the rapid halving ensures that the total cost is dominated only by the first full-resolution layer.

\paragraph{Super tokens as a queryable latent field.}
The pair $(\widebar{\bm{H}},\widebar{\bm{X}})$ forms a sparse, queryable latent representation of the current particle system.
Each super token aggregates local features, velocities, and material information from its represented particle group into a latent feature anchored at the corresponding position $\widebar{\bm{x}}_i$.
Unlike an implicit neural field stored entirely in network weights~\cite{crom}, or a neural eigenbasis precomputed for a specific shape family~\cite{shapespacespectra}, these super tokens are generated online from the current particle state at every timestep.
This distinction is the key to generalization:
within a trained dynamics category, the encoder can process held-out configurations, since super tokens are produced from whatever configuration is presented.
The decoder (Sec.~\ref{sec:decoder}) queries this representation by cross-attending from particle tokens to super tokens.

% ------------------------------------------------------------
\subsection{Super-Token Decoder}
\label{sec:decoder}
The decoder lifts coarse context back to particle resolution through $L_d$ layers.
The particle tokens $\bm{H}_{\mathrm d}^{(0)}\!=\!\{\bm{h}_i\}$ and positions $\tilde{\bm{X}}$ from the tokenizer are passed directly to the decoder, bypassing the encoder.
At decoder layer $\ell$, particle tokens query the super tokens through cross-attention~\cite{jaegle2021perceiver}:
\begin{align}
\widehat{\bm{H}}_\mathrm{d}^{(\ell)} &= \text{Cross-Attn}\!\bigl(\bm{H}_\mathrm{d}^{(\ell-1)},\, \tilde{\bm{X}};\; \widebar{\bm{H}},\, \widebar{\bm{X}}\bigr).
\label{eq:dec_ca}
\end{align}
Self-attention then refines the tokens:
\begin{align}
{\bm{H}}_\mathrm{d}^{(\ell)} = \text{Self-Attn}\!\bigl(\widehat{\bm{H}}_\mathrm{d}^{(\ell)},\, \tilde{\bm{X}}\bigr), \quad 
\bm{H}_\mathrm{d}^{(\ell)} \leftarrow \mathrm{MLP}\bigl(\bm{H}_\mathrm{d}^{(\ell)}\bigr). 
\label{eq:dec_sa}
\end{align}
All attention blocks use 3D RoPE~\cite{rope}, residual connections, and feed-forward networks.
Thus the decoder performs a learned coarse-to-fine lifting: the attention weights act as data-adaptive interpolation weights from super-token anchors to particles.
This resembles coarse-to-fine reconstruction in multilevel methods and reduced-order models~\cite{fulton2019latent,crom,shapespacespectra}, but the weights here are feature-dependent and recomputed for every timestep rather than fixed by geometry or a precomputed basis.
A final particle-wise MLP maps $\bm{H}_\mathrm{d}^{(L_d)}$ to $(\Delta\bm{x}_i,\Delta\bm{v}_i)$.
Full specifications are provided in the appendix.

\input{tabs/algorithm}
% ------------------------------------------------------------
\subsection{Training}
\label{sec:training}

We use the same architecture across all dynamics categories and train one network per category.
For a rollout window of $W$ states, including the initial ground-truth state $(W{\le}5)$, we start from $(\bm{X}_t,\bm{V}_t)$ and unroll the prediction-correction update autoregressively for $W{-}1$ steps, feeding each predicted state back as input.
The predicted states
$\{(\widehat{\bm{X}}_{t+n\Delta t},\widehat{\bm{V}}_{t+n\Delta t})\}_{n=1}^{W{-}1}$
are supervised against the ground-truth states
$\{(\bm{X}_{t+n\Delta t},\bm{V}_{t+n\Delta t})\}_{n=1}^{W{-}1}$.
The training loss is:
\begin{align}
\mathcal{L}
&=
\mathcal{L}_{\rm roll}
+
\lambda_{\rm phys}\,\mathcal{L}_{\rm phys},
\label{eq:total_loss}
\end{align}
where $\mathcal{L}_{\rm roll}$ is the mean per-particle, per-step position and velocity error averaged over the rollout, and $\mathcal{L}_{\rm phys}$ optionally supervises domain-specific physical constraints, e.g., $\mathcal{L}_{\rm phys}=\mathcal{L}_{\rm div}$, a divergence-free regularizer for incompressible fluids. Training is conducted on 8 NVIDIA A100
GPUs (80 GB VRAM per GPU).
% , with FlashAttention-2~\cite{dao2023flashattention2} to accelerate attention computation. 
The model has 156.73M parameters.
Optimization details are provided in the appendix.
% \lw{Here we use $\{(\bm{X}_{t+n\Delta t},\bm{V}_{t+n\Delta t})\}_{n=1}^{W{-}1}$ or $\{(\bm{X}_{t},\bm{V}_{t})\}_{t=1}^{W{-}1}$,and also needs to aline with algorithm}
% ~\zrpan{, where $n=1,\cdots,L$? Make clear because your layer label is also $L$ (recommend symbol change)}. The training objective is:

%% file: tabs/algorithm.tex
\begin{algorithm}[t]
\caption{Predictor-corrector transformer rollout}
\label{alg:rollout_our_model}
\begin{algorithmic}[1]
\REQUIRE Rollout window $W$, timestep $\Delta t$, mass matrix $\bm{M}$
\REQUIRE Initial state $(\bm{X}_t,\bm{V}_t)$, \textcolor[rgb]{0.9216,0.3804,0.0000}{\textbf{particle attributes}} $\bm{C}$, \textcolor[rgb]{0.1216,0.5451,0.5529}{\textbf{external forces}} $\{\bm{F}_{t+n\Delta t}\}_{n=0}^{W-2}$, rest-shape topology $\mathcal{T}$, boundary particles $(\bm{X}^{\textsc{b}},\bm{C}^{\textsc{b}})$
\ENSURE Predicted trajectory $\{({\bm{X}}_{t+n\Delta t},{\bm{V}}_{t+n\Delta t})\}_{n=1}^{W-1}$
\FOR{$n=0$ \TO $W-2$}
    \STATE \textit{// Prediction step (Eq.~\ref{eq:predictor})}
    \STATE $\tilde{\bm{V}}\leftarrow \bm{V}_{t+n\Delta t}+\Delta t\,\bm{M}^{-1}\bm{F}_{t+n\Delta t}$, \quad $\tilde{\bm{X}}\leftarrow \bm{X}_{t+n\Delta t}+\frac{\Delta t}{2}\!\left(\bm{V}_{t+n\Delta t}+\tilde{\bm{V}}\right)$
    \STATE \textit{// Correction step (Eq.~\ref{eq:corrector})}
    \STATE Build neighborhoods $\mathcal{N}^{\textsc{s}}$, $\mathcal{N}^{\textsc{b}}$, $\mathcal{N}^{\textsc{t}}$ from $(\tilde{\bm{X}},\,\bm{X}^{\textsc{b}},\,\mathcal{T})$
    \STATE $\bm{H} \leftarrow \textcolor[rgb]{0.5490,0.3843,0.2235}{\textbf{ParticleTokenizer}}(\tilde{\bm{X}},\tilde{\bm{V}},\bm{C},\mathcal{N}^{\textsc{s}},\mathcal{N}^{\textsc{b}},\mathcal{N}^{\textsc{t}},\bm{C}^{\textsc{b}})$
    \STATE $(\widebar{\bm{H}},\,\widebar{\bm{X}})\leftarrow \textcolor[rgb]{0.7843,0.3294,0.5412}{\textbf{SuperTokenEncoder}}(\bm{H},\,\tilde{\bm{X}})$
    \STATE $(\Delta\bm{X},\,\Delta\bm{V})\leftarrow \textcolor[rgb]{0.2471,0.4078,0.1490}{\textbf{SuperTokenDecoder}}(\bm{H},\,\tilde{\bm{X}},\,\widebar{\bm{H}},\,\widebar{\bm{X}})$
    \STATE \textit{// State update}
    \STATE ${\bm{X}}_{t+(n+1)\Delta t}\leftarrow \tilde{\bm{X}}+\Delta\bm{X}$, \quad ${\bm{V}}_{t+(n+1)\Delta t}\leftarrow \tilde{\bm{V}}+\Delta\bm{V}$
    % \STATE $\bm{X}_{t+(n+1)\Delta t}\leftarrow \widehat{\bm{X}}_{t+(n+1)\Delta t}$, \quad $\bm{V}_{t+(n+1)\Delta t}\leftarrow \widehat{\bm{V}}_{t+(n+1)\Delta t}$
\ENDFOR
\RETURN $\{({\bm{X}}_{t+n\Delta t},{\bm{V}}_{t+n\Delta t})\}_{n=1}^{W-1}$
\end{algorithmic}
\end{algorithm}

%% file: 4_experiments_v2.tex
\section{Experiments}
\label{sec_4}
\input{figs/main/fig_main}
We evaluate our model on six categories of particle dynamics, comparing it against state-of-the-art neural simulators and ablating key architectural design choices; model and training details are provided in the appendix.
For each dynamics category, we train a separate model using data generated by established solvers: Newtonian fluids from DFSPH~\cite{DFSPH}, cloth from GIPC~\cite{stiffGIPC,GIPC}, granular sand and non-Newtonian fluids from MPM~\cite{stomakhin2013material}, elastic shapes from VBD~\cite{vbd}, and protein molecular dynamics from OpenMM~\cite{eastman2023openmm8moleculardynamics}.
Within each category, we randomize material parameters, initial conditions, and boundary configurations across training sequences, and evaluate on held-out test scenes that vary these factors. 
Specifically, fluid scenes vary viscosity and obstacle placement; cloth scenes vary stiffness and density under sphere collisions; sand scenes vary initial pile shape and friction; non-Newtonian scenes vary rheological properties and slope angle; elastic solid scenes vary initial orientation on a fixed slope; and protein scenes vary initial atomic conformations.
Representative rollouts across all categories are shown in Fig.~\ref{fig:teaser},~\ref{fig:main}, and~\ref{fig:protein}, illustrating predicted results under varying materials, initial conditions, and boundary configurations. Quantitative results are reported in Table~\ref{tab:quan}, containing rollout mean-squared errors (MSE) for particle positions and velocities, averaged over all particles, frames, and test sequences, together with the degrees of freedom (DOF) and per-step compute cost.

\input{tabs/quantitative}

% The results show that our model achieves consistently low pair errors across diverse tasks, indicating accurate next-step prediction. As expected for neural simulators, rollout errors are higher than pair errors due to temporal accumulation, but the growth remains controlled on most scenarios.
% Combined with the qualitative results, this indicates that our model effectively captures particle-article and particle-boundary interactions, thereby mitigating long-horizon error accumulation and maintaining stable, physically plausible dynamics.

% Table~\ref{tab:val_pos_vel_mse} reports per-particle mean-squared errors(MSE) for position and velocity, evaluated on held-out test sequences in both one-step (pair) and auto-regressive (rollout) settings. The model achieves consistently low pair errors across all domains. Rollout errors are higher due to temporal accumulation, as is standard for auto-regressive neural simulators, but growth remains controlled over hundreds of frames.

% Notably, the model generalizes to unseen material parameters (stiffness, viscosity, friction), unseen initial conditions (object identity, orientation, velocity), and unseen boundary configurations (container geometry, obstacle pose) within each category.

\subsection{Comparison with Baselines}

We compare against GNS~\cite{gns}, DeepLagrangian~\cite{deepLagrange}, and Neural Operator~\cite{viswanath2024reduced,fno} on fluid, cloth, and sand tasks. As shown in Fig.~\ref{fig:baseline}, all baselines exhibit artifacts such as floating sand particles, explosive cloth behavior, and interpenetration for long rollouts, whereas our model maintains stable simulations that closely match the ground truth.

We attribute the improvement to three design choices. First, the prediction-correction decomposition factors out known external forces, so the learned corrector handles only inter-particle interactions. This separation also enables generalization to unseen force configurations at inference time. Second, the particle tokenizer captures fine-grained local interactions, while the super-token encoder and decoder provide global communication through self-attention and cross-attention; baselines typically offer one scale, requiring either many message-passing steps for long-range effects or sacrificing local detail. Third, progressive token merging supplies multigrid-like multi-resolution processing, resolving both fine contact details and long-range pressure or stress coupling simultaneously.

% Tab.~\ref{tab:baseline} reports quantitative errors, demonstrating that our model achieves the lowest errors across all three tasks.
% \input{tabs/tab_quantitative.tex}

% We also conduct comparisons on cloth, fluid and sand physics against several baselines~\cite{deepLagrange,gns,meshgraphnets,viswanath2024reduced}, with quantitative results reported in Table~\ref{tab:baseline}.
% Representative rollout sequences for all tasks are provided in Fig.~\ref{fig:baseline} and appendixary. \lw{TBD: Analysis of other methods.} Our model consistently outperforms competing baselines, achieving the lowest errors and demonstrating robust generalization while maintaining detailed particle interactions.

% \input{tabs/baseline}

% TODO: add baseline comparison figure (fig:baseline_comparison) and table (tab:baseline)
\subsection{Generalization}
We evaluate generalization along four axes: unseen initial and boundary conditions, long rollouts beyond the training horizon, unseen geometries, and unseen particle sampling densities. The architecture is designed to support: super tokens are generated from the current particle state at every timestep; the tokenizer uses relative displacements, making the model translation-equivariant; material parameters enter as per-particle attributes; and the prediction-correction separation factors out known external forces so the corrector learns transferable interaction patterns.

\paragraph{Initial and boundary conditions.}
Fig.~\ref{fig:deep_lagrange_dataset} shows two sequences from the dataset of~\cite{deepLagrange}. The first row is a training sequence whose initial fluid configuration is the closest match to the test case in the second row; despite this proximity, the fluid volume, container geometry, and obstacle placement all differ substantially (highlighted in orange). 
Our 800-frame rollout on the test case remains stable and closely tracks the ground truth, capturing splash patterns, fluid-obstacle interaction, and the long-term settling behavior, demonstrating generalization capabilities to unseen boundary and initial conditions.

\paragraph{Long rollouts.}
Fig.~\ref{fig:long_rollout} shows a cloth-twisting experiment where boundary vertices are driven by external torques of varying magnitudes. The model was trained on 200-frame sequences but tested here at 250 frames, requiring extrapolation beyond its training horizon. The cloth continues to accumulate twist without diverging, collapsing, or developing interpenetration, producing coherent winding motion. Long-horizon stability is a known difficulty for auto-regressive neural simulators as errors compound; 
the prediction-correction decomposition helps by providing a consistent intermediate state from the predictor before the corrector is applied.

\paragraph{Geometries.}
As shown in Fig.~\ref{fig:main}, the tiger-shaped sand pile never appeared in the training set, which contains only simple shapes such as cylinders and boxes, yet our model stably predicts its collapse with plausible granular flow and pile spreading. This suggests the model learns local interaction rules rather than memorizing shape-specific trajectories, so it transfers to novel geometries whose local particle neighborhoods remain within the training distribution. The same observation holds for the cloth category, where test meshes with unseen vertex counts and aspect ratios produce stable rollouts.

\paragraph{Particle sampling density.}
Fig.~\ref{fig:more_sampling} tests the fluid model, trained on approximately 7{,}000 particles, with altered initial arrangements and increased particle counts at higher resolutions. Rollouts remain stable across all tested densities. This robustness follows from the use of relative displacements in the tokenizer, which depend on inter-particle distance, allowing the model to process denser or sparser configurations than seen during training.

\subsection{Ablation Studies}
The ablations below isolate two core design choices: the interplay between local tokenization and global super-token communication, and the physics-informed regularization loss.

% We isolate the two main architectural design choices and evaluate the effect of physical regularization.

% \paragraph{Local vs.\ global communication.}
% We define three model variants. The \emph{full model} uses the complete pipeline: particle tokenizer, super-token encoder, and super-token decoder. The \emph{w/o local} variant removes the tokenizer's neighborhood aggregation branches; particle tokens are built only from per-particle features through a pointwise MLP, so the model has no explicit local interaction encoding. The \emph{w/o global} variant keeps the particle tokenizer but removes the super-token encoder and decoder entirely; predictions are made from local particle tokens alone using a pointwise MLP decoder.

% Table~\ref{tab:ablation_local_global} reports rollout errors across fluid, cloth, and sand. Removing local tokenization degrades short-range interaction modeling: boundary penetration increases, cloth-sphere contact quality deteriorates, and fine-scale granular structure is lost. Removing the super-token pathway degrades long-range coordination: fluid pressure waves dissipate too quickly, cloth deformation far from the contact zone becomes inconsistent, and the spatial extent of sand collapse is underestimated. The full model is needed because particle dynamics require both local physical supports and global communication.

\begin{figure}
    \centering
    \includegraphics[width=\linewidth]{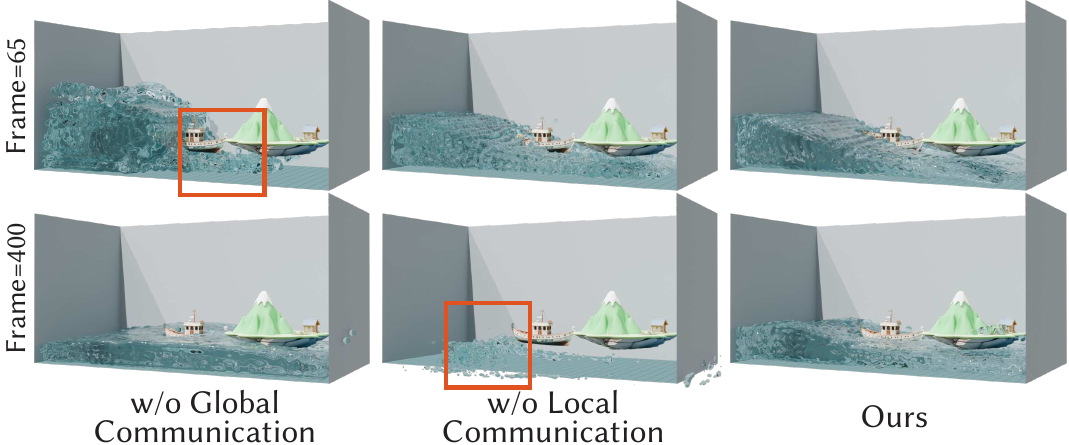}
    \vspace*{-2em}
    \caption{\textbf{Ablation on local and global communication.} Three variants on the fluid task: w/o the super-token encoder-decoder (local only), w/o the tokenizer's neighborhood branches (global only), and the full model. Removing either component degrades rollout quality.}
    \label{fig:local_global_ab}
\end{figure}

% TODO: add local/global ablation table (tab:ablation_local_global)
\paragraph{Local and global communication.}
We examine the effect of local and global communication on the fluid task. The \emph{local-only} model removes the super-token encoder and decoder and predicts the correction quantity from particle tokens alone, reducing to the Deep Lagrangian approach~\cite{deepLagrange} without multi-level information exchange. The \emph{global-only} model removes the neighborhood aggregation branches in the tokenizer, so each token encodes only the single particle under consideration rather than summarizing information from a local neighborhood, and embeds positions and velocities directly into token space. As shown in Fig.~\ref{fig:local_global_ab}, the local-only model fails to propagate long-range pressure information, while the global-only model loses fine-grained particle interactions and accumulates error over time. The full model, combining both pathways, produces the most accurate rollouts.
% As demonstrated by Fig.\ref{fig:baseline_comparison}, neither side alone is sufficient: the local-only model fails to capture correct dynamics, and the global-only model, built by removing particle interactions in our Particle Tokenizer (Sec.~\ref{sec:tokenizer}) and embedding positions and velocities directly into token space, struggles to capture fine-grained particle dynamics and accumulates errors over time. In contrast, our model integrates both local and global communication, producing stable and accurate long-range rollouts across all test cases.

\begin{wraptable}{r}{0.48\linewidth}
\vspace{-1.2em}
\centering
\setlength{\tabcolsep}{4pt}
% \caption{Effect of divergence-free regularization on fluid rollout errors.}
% \label{tab:ablation_div}
\begin{tabular}{ccc}
\toprule
\textbf{Metric} & \textbf{w/o} $\mathcal{L}_{\rm div}$ & \textbf{w/} $\mathcal{L}_{\rm div}$ \\
\midrule
\textbf{Pos. Loss} & 0.517 & 0.535 \\
\textbf{Vel. Loss} & 0.263 & 0.292 \\
$\mathcal{L}_{\rm div}$ & 0.015 & 0.008 \\
\bottomrule
\end{tabular}
\vspace{-0.8em}
\end{wraptable}
\paragraph{Physical regularization.}
We test the effect of adding a divergence-free regularizer $\mathcal{L}_{\rm div}$ to $\mathcal{L}_{\rm phys}$ on the fluid task (Fig.~\ref{fig:main}). The divergence is computed using the standard SPH estimator with the same kernel used by the DFSPH reference solver~\cite{DFSPH}, applied to the predicted particle positions and velocities at each rollout step. The inset table reports rollout position and velocity MSE alongside the divergence residual. Adding $\mathcal{L}_{\rm div}$ reduces the divergence error while maintaining comparable trajectory accuracy.
We also ablate network hyperparameters and architectural choices on the cloth task (Fig.~\ref{fig:main}); full results are reported in the appendix.

\subsection{Applications}
Our architecture unlocks various downstream applications: the predictor's explicit external-force integration (Eq.~\ref{eq:predictor}) enables generalization to unseen force configurations for interactive control; the differentiable neural corrector enables gradient-based optimization of physical parameters for inverse design; and the generic particle representation allows training on real-world point-cloud observations. Fast single-GPU inference makes these applications practical.

% Leveraging the key properties of our model, which combines accuracy, efficiency, and a unified representation for particle-based dynamics, we explore a variety of applications to demonstrate its range of tasks and practical utility.

\paragraph{Interactive Control}
% In the particle tokenizer (Sec.~\ref{sec:tokenizer}), the model first computes a physics-informed intermediate state that explicitly integrates the known external forces. In this way, the model can generalize to physics with instantaneous external force applied. Leveraging this capability along with the model high inference speed, we enable interactive control applications.

% As illustrated in Fig.~\ref{fig:control}, we implement an interactive system using Polyscope~\cite{polyscope}, where users can select different loaded meshes and specify the direction, region, and magnitude of instantaneous forces applied. Our model demonstrates strong generalization to external forces, producing stable and physically plausible simulations even for force configurations not seen during training.
% In the particle tokenizer (Sec.~\ref{sec:tokenizer}), the model computes a physics-informed intermediate state that explicitly incorporates known external forces, enabling generalization to physics with instantaneous force application. Combined with high inference speed, this allows for interactive control applications. As shown in Fig.~\ref{fig:control}, we build an interactive system using Polyscope~\cite{polyscope}, where users can select meshes and specify the direction, location, and magnitude of applied forces. The model generalizes robustly to unseen force configurations, producing stable and physically plausible simulations.
Because the prediction step explicitly integrates external forces, the model generalizes to force configurations not seen during training. Combined with fast inference, this enables interactive applications. As shown in Fig.~\ref{fig:control}, we build a real-time interface using Polyscope~\cite{polyscope} where users apply forces of arbitrary direction, location, and magnitude.

% \begin{figure}
%     \centering
%     \includegraphics[width=0.48\textwidth]{figs/control.pdf}
%     % \caption{\textbf{Interactive force control on deformable objects.} The left panel shows user-specified force parameters, including the region of application, direction, and duration, while the three examples on the right illustrate the corresponding controllable motion responses generated by our model in real time.}
%     \caption{\textbf{Interactive force control.} Left: user-specified force parameters (application region, direction, and duration). Right: three examples of the resulting model rollouts under forces not seen during training.}
%     \label{fig:control}
% \end{figure}

\paragraph{Inverse Design}
% Because the corrector $\mathcal{F}_\theta$ is differentiable through the neural computations for fixed neighborhoods and token assignments, we can optimize physical parameters by backpropagating through auto-regressive rollouts. We demonstrate this on a friction-estimation task: a deformable object is released on a slope, and the goal is to recover the friction coefficient $\mu$ that places the object at a target stop location $\mathbf{x}^\star$. We freeze the trained simulator and minimize the design objective $\mathcal{J}(\mu) = \|\mathbf{s}(\mu) - \mathbf{x}^\star\|^2$ with respect to $\mu$ using gradient descent through the unrolled trajectory $\mathbf{s}(\mu)$.

% Our model can stably simulate highly complex interactions between objects and their environment.

% Importantly, the framework is unified, enabling inverse design across a wide range of dynamical systems. Its particle-based representation further extends the potential to tackle real-world design problems.
Since gradients can be backpropagated through the neural rollout, we can optimize physical parameters by backpropagating through auto-regressive rollouts. We consider a setting where a duck-shaped inflatable mesh slides down a ramp and glides along the ground; the goal is to recover the friction coefficient $\mu$ that brings the duck to rest at a prescribed target location. We freeze the trained model and optimize $\mu$ alone via gradient descent through the rollout. Fig.~\ref{fig:inverse} compares trajectories under the initial and optimized friction values alongside the optimization curve: the recovered $\mu$ places the duck at the target, and the loss converges smoothly.

\paragraph{Learning from Real-World Data}
Since our model operates on particle representations, we can train it directly on real-world point cloud observations without a classical simulator. Fig.~\ref{fig:real_world} shows three manipulation sequences from PhysTwin~\cite{jiang2025phystwin}, comparing real-world observations with our model's predicted trajectories. The predicted rollouts qualitatively match the observed motion.

%% file: figs/main/fig_main.tex
\begin{figure*}[t]
  \centering
  \includegraphics[width=1.0\linewidth]{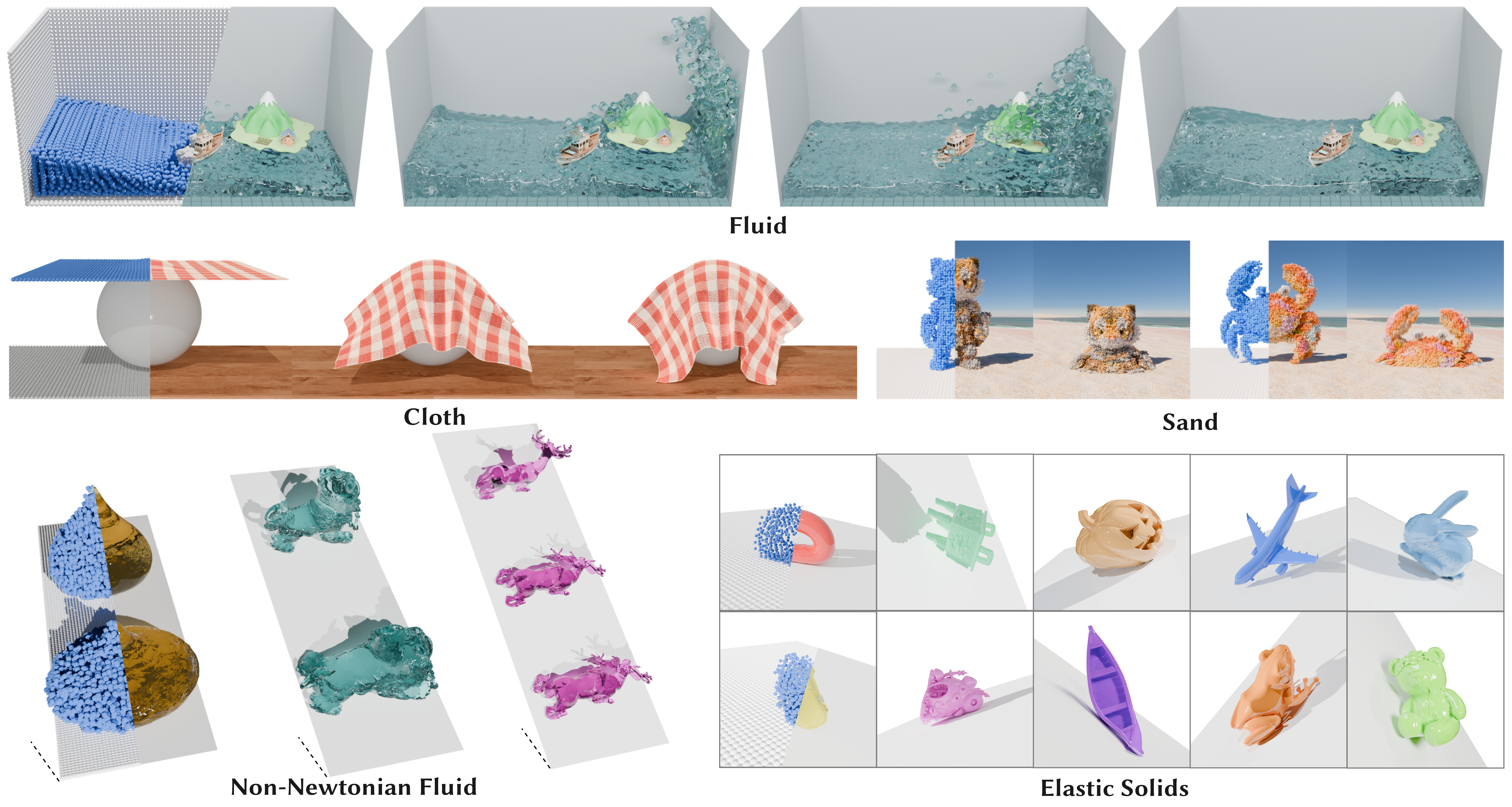}
  \vspace*{-2em}
    \caption{\textbf{Qualitative results across five of six dynamics categories: Newtonian fluids, cloth, granular sand, non-Newtonian fluids, and elastic solids.} 
    Example sequences show Newtonian fluids flowing with varying viscosity and obstacle placements in a container, cloth colliding with a sphere, granular sand collapsing with varied initial shapes and friction, non-Newtonian fluids with varying rheological properties on different slopes, and elastic solids with varying initial rotation on a fixed slope. 
    All test sequences use unseen configurations. In the first figure of each sequence, blue spheres indicate predicted particles, and white spheres indicate boundary samples. 
    Results for protein are shown in Fig.~\ref{fig:protein}.
    % \lw{name in figs should be aligned.}
    % \lw{For more description of unseen target.}
  % \md{This figure needs more design. (1) The cloth thing is not beautiful enough to show a full sequence, but it does not show any ability to control material properties like stiffness as suggested by the text. (2) Similar to cloth, it's just a bear bouncing, not a beautiful video. It also does not compare anything. (3)(4) Are they equivalent? Where is non-Newtonian? And what is the expected behavior? (5) Are left and right parts equivalent? Is there a comparison against GT? I think fluid/sand is more beautiful, if we choose to display one sequence, maybe use the fluid.}
  }
  \label{fig:main}
\end{figure*}

%% file: tabs/quantitative.tex
\begin{table}[t]
\centering
\setlength{\tabcolsep}{2.5pt}
\small
\caption{\textbf{Accuracy and inference cost on test sets.}
Auto-regressive rollout MSE for particle positions and velocities, averaged over particles, frames, and test sequences. DOF: degrees of freedom. Throughput is reported in frames/s; per-step compute cost is in TFLOPs. All measurements are conducted on a single NVIDIA A100 GPU (80 GB VRAM).}
\label{tab:quan}
\vspace{-1em}
\begin{tabular}{@{}c cccccc@{}}
\toprule
\textbf{Metric} & \textbf{Fluid} & \textbf{Cloth} & \textbf{Sand} & \textbf{Non-Newt.} & \textbf{Elastic} & \textbf{Protein} \\
\midrule
\textbf{\# of Frames} & 600 & 90 & 100 & 250 & 100 & 100 \\
\textbf{DOFs}    & 42{,}120 & 29{,}400 & 60{,}000 & 60{,}000 & 52{,}542 & 61{,}884 \\
\midrule
\textbf{Pos. Error} {\scriptsize($\times 10^{-3}$)} & 516.585 & 0.077 & 0.054 & 0.706 & 0.158 & 0.001 \\
\textbf{Vel. Error} {\scriptsize($\times 10^{-3}$)} & 263.149 & 0.454 & 1.442 & 1.760 & 3.530 & 0.092 \\
\midrule
\textbf{Speed} (frames/s) & 16.176 & 24.050 & 11.500 & 11.168 & 11.708 & 14.695 \\
\textbf{Cost} (TFLOP/step) & 1.323 & 0.923 & 1.886 & 1.886 & 1.651 & 0.944 \\
\bottomrule
\end{tabular}
\end{table}

%% file: 7_conclusion.tex
\section{Conclusion and Discussion}
We presented a prediction-correction particle transformer for Lagrangian dynamics, trained on cloth, elastic solids, Newtonian and non-Newtonian fluids, granular materials, and molecular dynamics. The architecture combines local particle tokenization with a super-token encoder-decoder, capturing both local interactions and global coupling within a single fixed architecture. The key design choice is that super tokens are constructed from the full dynamical state at every timestep, making the global communication input-dependent rather than tied to a fixed basis or geometry. The model generalizes to unseen materials, boundary configurations, initial conditions, and external forces, and supports downstream tasks including interactive control, inverse design, and learning from real-world observations. 
% Across six dynamics categories, it achieves lower rollout errors than existing neural simulators while covering a broader range of physical phenomena within one architecture.

Several directions remain for future work. The main limitation is memory: the first encoder layer applies self-attention at full particle resolution, currently supporting up to approximately 50{,}000 particles on a single 80GB A100 GPU. Memory-efficient attention variants such as sparse attention~\cite{child2019generating} could extend this budget to larger scenes. 
We currently train one parameter set per dynamics category; training a single model with shared weights across all categories would move from a unified architecture to a unified model, though scaling to the combined data volume and maintaining per-category accuracy will pose additional challenges. 
Extending the per-particle attribute to include quantities such as temperature or strain history would broaden the range of addressable phenomena, and tighter integration with differentiable reconstruction pipelines~\cite{hafner2023mastering} could further close the gap between learned simulation and real-world deployment.

%% file: extra_figs_v2.tex
% ============================================================
% Extra Figure Page 1: Breadth, Generalization, Inverse Design
% ============================================================
\newpage

\begin{minipage}{\linewidth}
\centering
\includegraphics[width=\linewidth]{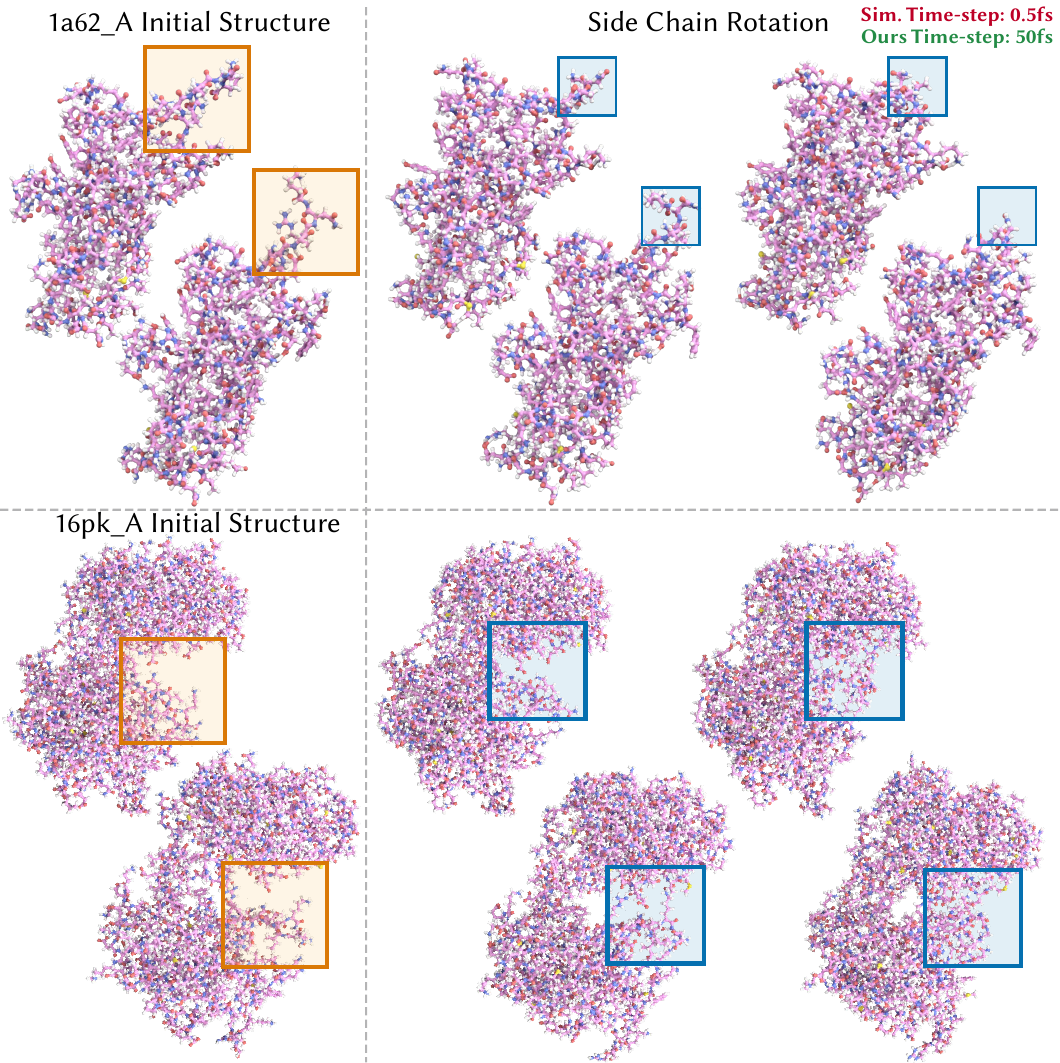}
\vspace*{-2em}
\captionof{figure}{\textbf{Side-chain rotation prediction on proteins 1a62\_A and 16pk\_A.} Starting from different initial conformations (orange), the model predicts side-chain rotational dynamics (blue) at 50\,fs timesteps, 100$\times$ larger than the 0.5\,fs steps used in traditional simulations.}
\label{fig:protein}
\end{minipage}

\vspace{1.6em}

\begin{minipage}{\linewidth}
\centering
\includegraphics[width=0.9\linewidth]{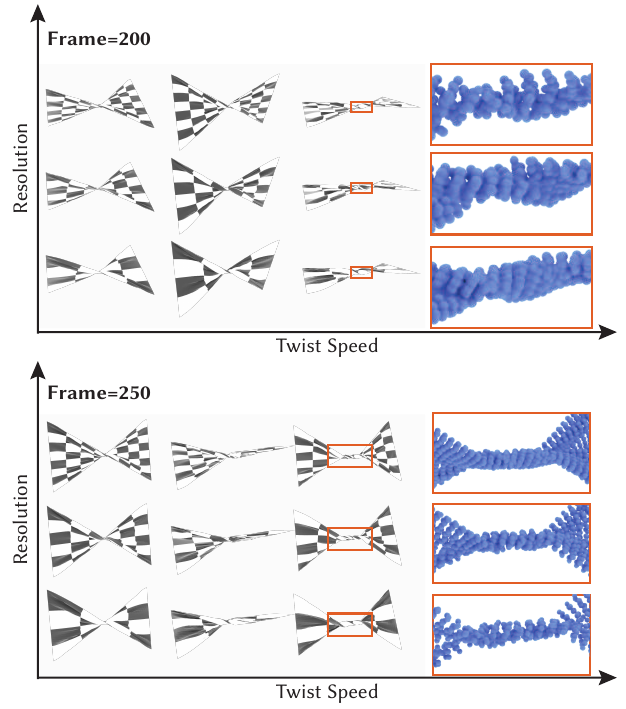}
\vspace*{-1em}
\captionof{figure}{\textbf{Long rollouts and unseen actuation.} Cloth-twisting sequences under actuation strengths not seen during training. The model remains stable up to 250 frames, beyond the 200-frame training horizon.}
\label{fig:long_rollout}
\end{minipage}

\begin{minipage}{\linewidth}
\centering
\includegraphics[width=1.0\textwidth]{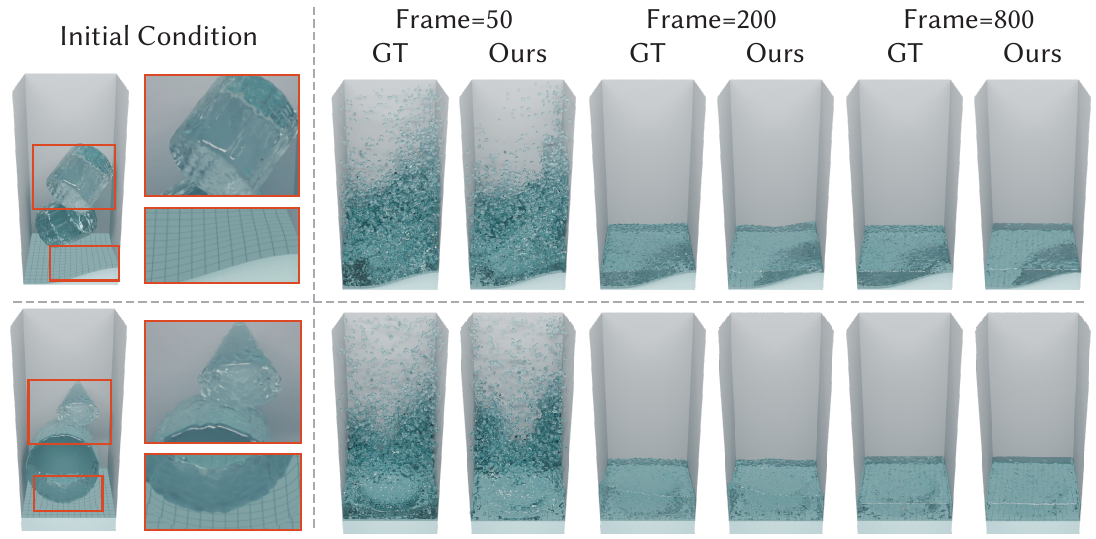}
\vspace*{-2em}
\captionof{figure}{\textbf{Generalization to unseen initial and boundary conditions.} First row: the training sequence whose initial configuration is closest to the test case in the second row. Orange boxes highlight differences in fluid volume and container geometry. Despite substantial differences, the model produces a stable 800-frame rollout that closely tracks the ground truth.
}
\label{fig:deep_lagrange_dataset}
\end{minipage}
\vspace{0.8em}

\begin{minipage}{\linewidth}
\centering
\includegraphics[width=\linewidth]{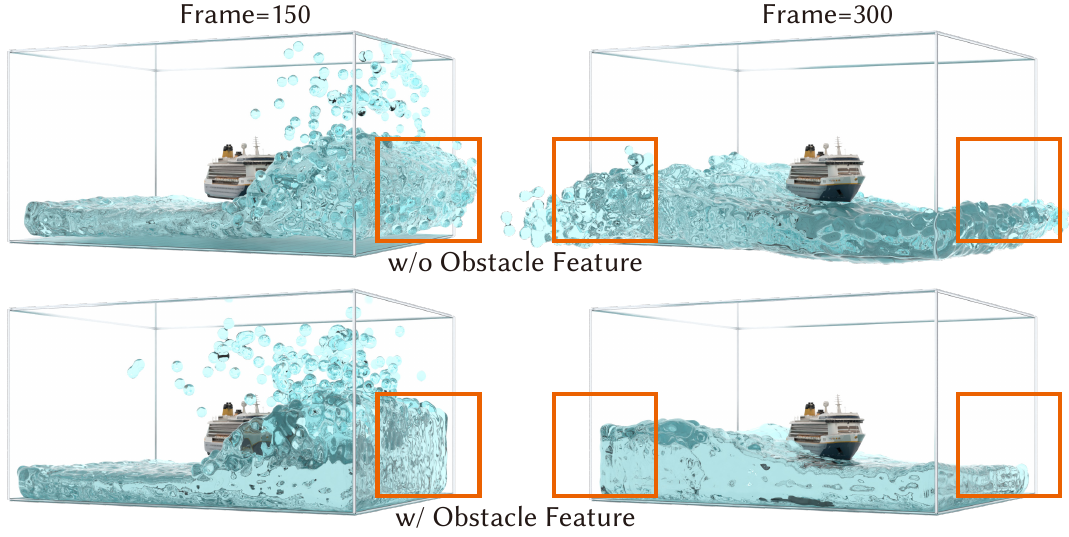}
\vspace*{-2em}
\captionof{figure}{\textbf{Ablation on particle-boundary interaction.} Fluid rollouts without (top) and with (bottom) the boundary branch of the particle tokenizer. Without boundary features, the model produces penetration.}
\label{fig:baseline_comparison}
\end{minipage}

\begin{minipage}{\linewidth}
\centering
\includegraphics[width=\linewidth]{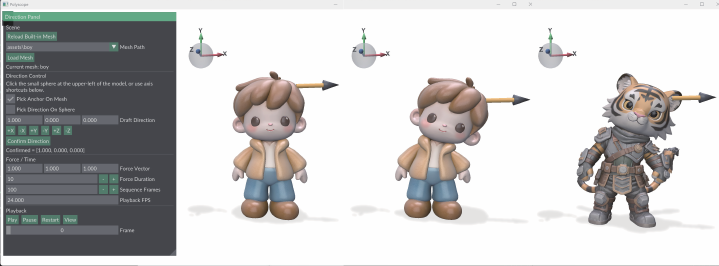}
\vspace*{-2em}
\captionof{figure}{\textbf{Interactive force control.} Left: user-specified force parameters (application region, direction, and duration). Right: three examples of the resulting model rollouts under forces not seen during training.}
\label{fig:control}
\end{minipage}

\begin{minipage}{\linewidth}
\centering
\includegraphics[width=\linewidth]{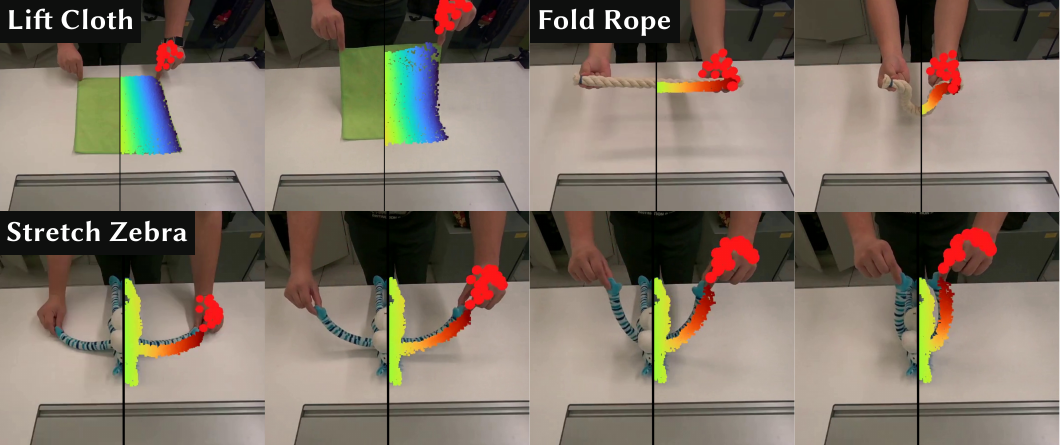}
\vspace*{-2em}
\captionof{figure}{\textbf{Real-world manipulation.} Predictions on three tasks from PhysTwin~\cite{jiang2025phystwin}: lifting cloth, stretching a zebra toy, and folding rope. Each subfigure: real-world observation (left) and predicted particle trajectories (right).}
\label{fig:real_world}
\end{minipage} 

% \begin{minipage}{\linewidth}
% \centering
% \includegraphics[width=\linewidth]{figs/gen_more_sampling.pdf}
% \vspace*{-2em}
% \captionof{figure}{\textbf{Generalization to particle sampling density.} The fluid model, trained on approximately 7k particles, produces stable rollouts under altered initial particle arrangements (7k) and increased particle counts (8k, 9k).}
% \label{fig:more_sampling}
% \end{minipage}

% \begin{minipage}{\linewidth}
% \centering
% \includegraphics[width=\linewidth]{figs/abl_obs.pdf}
% \vspace*{-2em}
% \captionof{figure}{\textbf{Ablation on particle-boundary interaction.} Fluid rollouts without (top) and with (bottom) the boundary branch of the particle tokenizer. Without boundary features, the model produce penetration.}
% \label{fig:baseline_comparison}
% \end{minipage}

% \begin{minipage}{\linewidth}
% \centering
% \includegraphics[width=\linewidth]{figs/control.pdf}
% \vspace*{-2em}
% \captionof{figure}{\textbf{Interactive force control.} Left: user-specified force parameters (application region, direction, and duration). Right: three examples of the resulting model rollouts under forces not seen during training.}
% \label{fig:control}
% \end{minipage}

\vspace{0.8em}

\begin{minipage}{\linewidth}
\centering
\includegraphics[width=\linewidth]{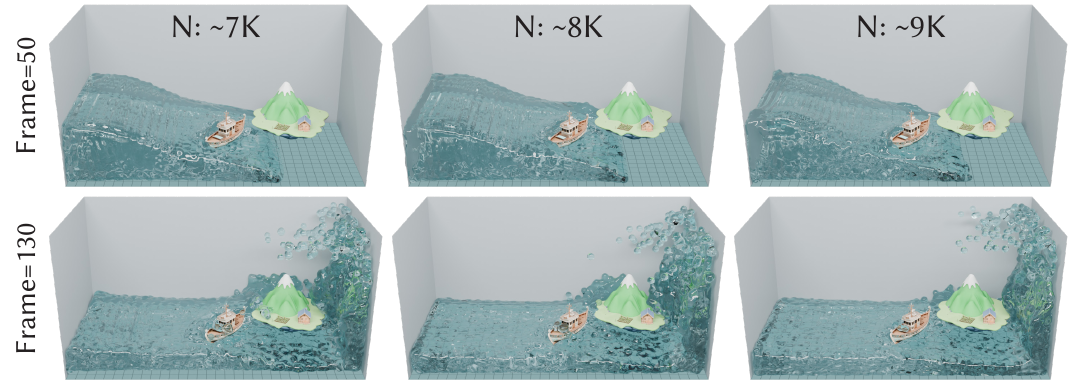}
\vspace*{-2em}
\captionof{figure}{\textbf{Generalization to particle sampling density.} The fluid model, trained on approximately 7k particles, produces stable rollouts under altered initial particle arrangements (7k) and increased particle counts (8k, 9k).}
\label{fig:more_sampling}
\end{minipage}

\vspace{1.6em}

\begin{minipage}{0.95\linewidth}
\centering
\includegraphics[width=\linewidth]{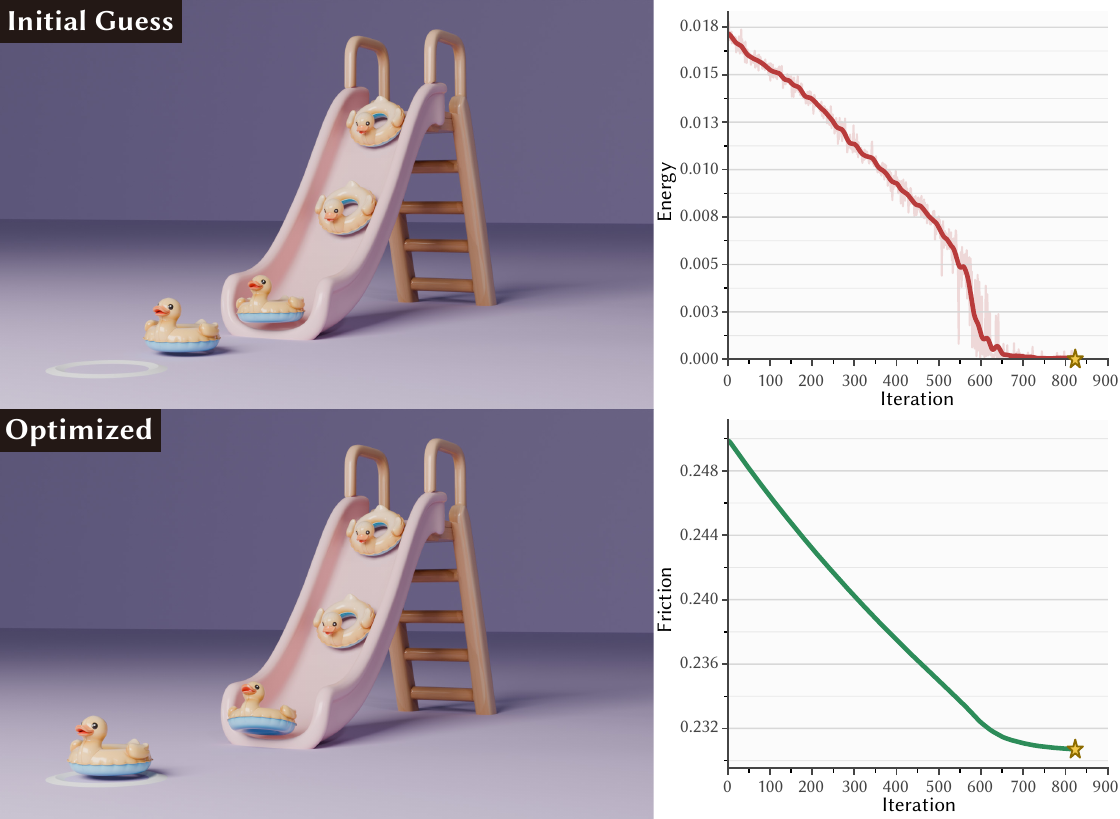}
\vspace*{-2em}
\captionof{figure}{\textbf{Inverse design.} Left: forward rollout under the initial friction guess (top) and the optimized friction (bottom). Right: design objective and recovered friction coefficient over optimization iterations.}
\label{fig:inverse}
\end{minipage}

\vspace{1.6em}

% \begin{minipage}{\linewidth}
% \centering
% \includegraphics[width=\linewidth]{figs/control.pdf}
% \vspace*{-2em}
% \captionof{figure}{\textbf{Interactive force control.} Left: user-specified force parameters (application region, direction, and duration). Right: three examples of the resulting model rollouts under forces not seen during training.}
% \label{fig:control}
% \end{minipage}

\begin{minipage}{\linewidth}
\centering
\includegraphics[width=\linewidth]{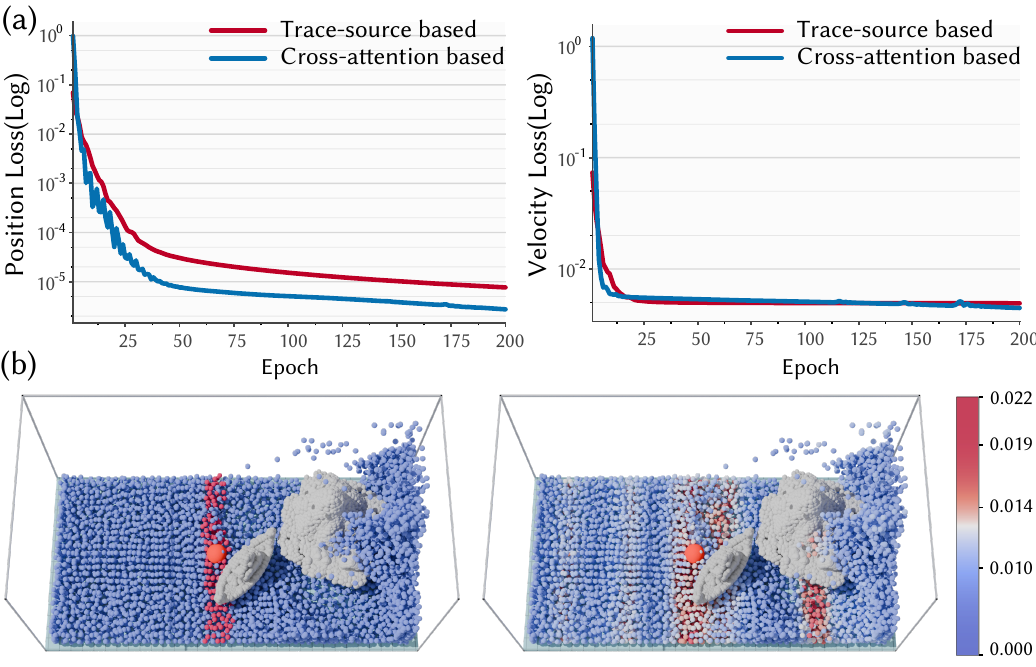}
\vspace*{-2em}
\captionof{figure}{\textbf{Cross-attention decoder vs.\ trace-source decoder.}
(a) Training loss of two diagnostic probes attached to the same frozen super-token encoder. The cross-attention probe achieves lower position and velocity errors than the trace-source probe.
(b) Influence maps for the two probes: warmer colors indicate stronger influence. The trace-source decoder shows primarily local support, while the cross-attention decoder captures non-local, motion-aware coupling.}
\label{fig:rom}
\end{minipage}
% ============================================================
% Extra Figure Page 2: Baselines, Applications, Analysis
% ============================================================
\newpage

\begin{minipage}{\linewidth}
\centering
\includegraphics[width=\linewidth]{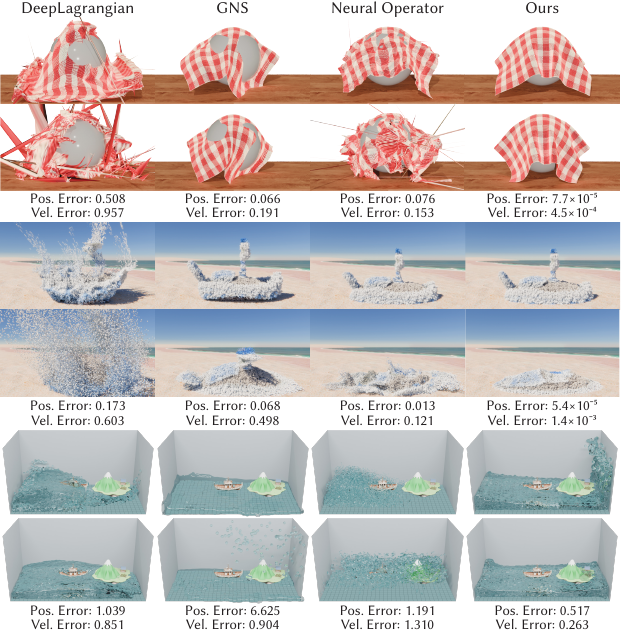}
\vspace*{-2em}
\captionof{figure}{\textbf{Baseline comparison on fluid, cloth, and sand.}
Rollout sequences for DeepLagrangian~\cite{deepLagrange}, GNS~\cite{gns}, Neural Operator~\cite{fno,viswanath2024reduced}, and our model, compared against ground truth (GT). Reported errors are auto-regressive rollout MSE for position and velocity, averaged over particles, frames, and test sequences.}
\label{fig:baseline}
\end{minipage}

\vspace{0.8em}

% \begin{minipage}{\linewidth}
% \centering
% \includegraphics[width=\linewidth]{figs/control.pdf}
% \vspace*{-2em}
% \captionof{figure}{\textbf{Interactive force control.} Left: user-specified force parameters (application region, direction, and duration). Right: three examples of the resulting model rollouts under forces not seen during training.}
% \label{fig:control}
% \end{minipage}

\vspace{0.8em}

\vspace{0.8em}

\vspace{0.8em}

\begin{minipage}{\linewidth}
\centering
\includegraphics[width=\linewidth]{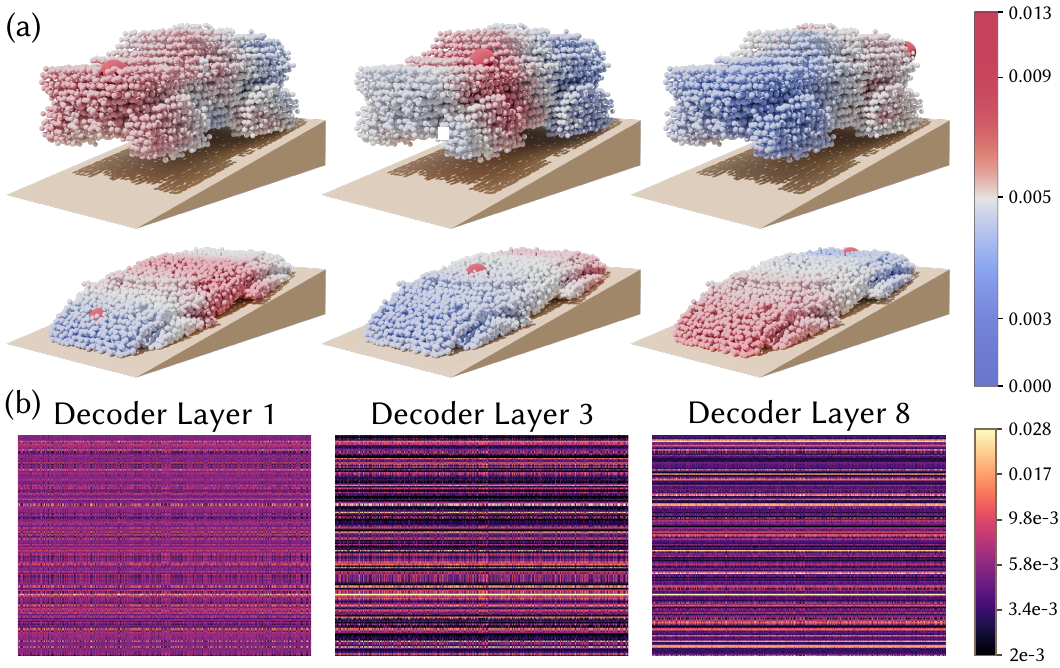}
\vspace*{-2em}
\captionof{figure}{\textbf{Decoder attention patterns.}
(a) Influence maps of selected super tokens on particles at two frames across 8 decoder layers; warmer colors indicate stronger influence. Each super token affects spatially distributed particles, reflecting global coupling.
(b) Cross-attention heat maps from decoder layers 1, 3, and 8 (rows: super tokens, columns: particles). Horizontal high-response bands indicate that super tokens induce globally coupled updates across many particles. Experiment uses the non-Newtonian model from Fig.~\ref{fig:main}.}
\label{fig:decoder_global}
\end{minipage}

\clearpage

%% file: Appendix_arxiv.tex
\section{Method Details}
\label{app:method_details}

\subsection{Learnable Kernel for Particle Tokenizer}
For branch $k\in\{\textsc{s},\textsc{b},\textsc{t}\}$, the kernel $\bm{W}_k$ is parameterized by a learnable 3D lattice
\[
\Theta_k \in \mathbb{R}^{G \times G \times G \times C_{\mathrm{in}}^k \times C_{\mathrm{out}}^k},
\]
where each vertex stores a $C_{\mathrm{in}}^k \!\times\! C_{\mathrm{out}}^k$ mixing matrix and $G$ is the lattice resolution per axis. Given relative displacement $\bm{r}^k_{ij}$ with $\|\bm{r}^k_{ij}\| \le R_k$ (zero otherwise), we normalize to grid coordinates:
\[
\hat{\bm{r}}^k_{ij} = \bm{r}^k_{ij} / R_k, \qquad
\xi_d = \tfrac{\hat{r}_d + 1}{2}(G-1), \quad d\in\{x,y,z\},
\]
and evaluate $\bm{W}_k$ by trilinear interpolation over the eight enclosing lattice vertices:
\begin{align}
\bm{W}_k(\bm{r}^k_{ij}) &= 
\sum_{\bm{\delta}\in\{0,1\}^3} 
\omega_{\bm{\delta}}(\bm{t}) \; \Theta_k[\bm{n}+\bm{\delta}], \\
\omega_{\bm{\delta}}(\bm{t}) &= 
\prod_{d\in\{x,y,z\}} \big(\delta_d\, t_d + (1{-}\delta_d)(1{-}t_d)\big),
\end{align}
where $\bm{n}=\lfloor\bm{\xi}\rfloor$, $\bm{t}=\bm{\xi}-\bm{n}$, and $\Theta_k[\bm{n}+\bm{\delta}]$ is the mixing matrix at lattice corner $\bm{n}+\bm{\delta}$. The resulting kernel is continuous and piecewise linear inside the support radius $R_k$. Each neighbor contributes $\bm{W}_k(\bm{r}^k_{ij})\,\bm{u}^k_{ij}$, and the branch output is $\bm{a}^k_i = \sum_{j\in\mathcal{N}^k_i} \bm{W}_k(\bm{r}^k_{ij})\,\bm{u}^k_{ij}$.

% \subsection{Final Particle-wise MLP}
\subsection{Model Specification}
The particle tokenizer uses three neighborhood branches with kernels on a $4\times4\times4$ learnable lattice at base width $384$. The spatial and topology-guided branches are concatenated as two $192$-channel streams, the boundary branch contributes $384$ channels, and the particle self-feature branch contributes $384$ channels, yielding $\bm{h}_i\in\mathbb{R}^{1152}$.

The super-token encoder applies $L_e=6$ layers of self-attention and token merging at width $1152$, with $16$ attention heads (head dimension $72$) and bipartite merging ($k=2$). 3D RoPE with rotary dimension $72$ is applied at every layer using current token coordinates.

The super-token decoder uses $L_d=8$ layers, each consisting of cross-attention (particle queries, super-token keys/values), particle self-attention, and a feed-forward network, at width $1152$ with $12$ heads (head dimension $96$), FFN hidden width $512$, and dropout $0.1$. 3D RoPE with rotary dimension $48$ is applied in both cross-attention and self-attention, using particle positions for queries and super-token positions for keys.

The final prediction head maps $\bm{H}_\mathrm{d}^{(L_d)}$ to $(\Delta\bm{x}_i,\Delta\bm{v}_i)\in\mathbb{R}^6$ through a five-layer MLP ($1152\!\to\!512\!\to\!512\!\to\!512\!\to\!512\!\to\!6$), where the first four layers use Linear+ReLU+LayerNorm and the last is linear without activation.

The parameter count by module: tokenizer $249{,}600$; super-token encoder $55{,}842{,}060$; super-token decoder $99{,}255{,}296$; prediction head $1{,}385{,}478$; total $156{,}732{,}450$.

\section{Experiment Details}
\label{app:exp_details}
\subsection{Training Details}
We train with AdamW~\cite{adamw} ($\beta_1=0.9$, $\beta_2=0.999$, weight decay $5\times10^{-4}$, base learning rate $1\times10^{-4}$). Training uses 8 NVIDIA A100 GPUs (80\,GB each) with FlashAttention-2~\cite{dao2023flashattention2}. The learning rate follows a linear warmup over the first 8{,}000 steps (start factor $0.01$), then cosine annealing over 400{,}000 steps to $5\times10^{-6}$. Batch size is 1; the number of epochs varies by category. One model is trained per dynamics category, shared across all systems within that category.

\subsection{Main Results}
\label{app:main}
\paragraph{Newtonian Fluids.}
We simulate dam-break dynamics in a container of size $4\times2\times2$ ($x \in [-2,\,2]$, $y,z \in [-1,\,1]$), where fluid interacts with two fixed rigid bodies: a boat and an island. Each sequence uses randomized fluid viscosity $\nu \in [4.0\times10^{-3},\,4.0\times10^{-2}]$ and is simulated for 600 frames at $\Delta t = 0.01$\,s using DFSPH~\cite{DFSPH} in SPlisHSPlasH~\cite{SPlisHSPlasH_Library}. The boat and island rotations around the $y$-axis are randomized; the boat centroid is sampled in $x,z \in [-0.3,\,0.3]$, $y \in [0.38,\,0.45]$; the island centroid is sampled in $x \in [0.8,\,1.2]$, $z \in [-0.3,\,0.3]$ with fixed $y=0.45$.

Per-particle attributes $\bm{C}$ consist of the normalized viscosity $\hat{\nu} = (\nu - \mu_\nu)/(\sigma_\nu + 10^{-8})$, where $\mu_\nu$ and $\sigma_\nu$ are the mean and standard deviation across sequences. Boundary particles $(\bm{X}^{\textsc{b}}, \bm{C}^{\textsc{b}})$ are sampled from the surfaces of the container walls (54{,}000 samples), boat (3{,}000), and island (3{,}000), with outward surface normals as boundary attributes $\bm{C}^{\textsc{b}}$. The dataset contains 200 sequences, split into 160 training, 20 validation, and 20 test.
% As illustrated in Fig.~\ref{fig:main}, Our model accurately reproduces fluid dynamics and interactions with the boat, island, and container walls, maintaining penetration-free behavior throughout long rollouts. The results demonstrate that the model maintains stable generalization across variations in simulation parameters and boundary configurations, specifically manifested through differing fluid viscosities and the poses of rigid obstacles.

\paragraph{Cloth.}
A triangular cloth mesh with 4{,}900 vertices falls onto a fixed sphere (radius $0.3$, center height $0.3$) above a ground plane at height $-0.01$, released from height $0.62$. For each sequence, Young's modulus $E\in[1.0\times10^4,\,2.0\times10^5]$ and density $\rho\in[100,\,400]$ are randomized. Stiff-GIPC~\cite{GIPC,stiffGIPC} generates 90-frame trajectories at $\Delta t=0.02$\,s.

Each cloth vertex is treated as a simulated particle. Per-particle attributes $\bm{C}$ consist of the normalized $(E, \rho)$ repeated across all vertices. The topology $\mathcal{T}$ is the initial triangular mesh connectivity. Boundary particles $(\bm{X}^{\textsc{b}}, \bm{C}^{\textsc{b}})$ are sampled from the sphere surface (5{,}000 samples) and ground plane (5{,}000), with outward normals as $\bm{C}^{\textsc{b}}$. The dataset contains 200 sequences, split into 160 training, 20 validation, and 20 test.

% As illustrated in Fig.~\ref{fig:main}, the model accurately reproduces the cloth sphere collision process, including contact formation, sliding, and post-impact settling. The results indicate that the model generalizes reliably to unseen material and contact conditions, specifically variations in stiffness, density, and friction coefficient.

\paragraph{Granular Sand.}
Sand piles are dropped onto a rigid ground plane, simulated with MPM~\cite{stomakhin2013material} in NVIDIA Newton~\cite{newton2025}. Each sequence is initialized from one of 20 predefined mesh geometries, with the interior voxel-sampled into 10{,}000 particles. The internal friction coefficient is randomized within $\mu \in [0.20,\,1.00]$; other parameters are fixed at density $\rho = 1000$, Young's modulus $E = 1.0\times10^{15}$, and Poisson's ratio $\nu = 0.3$. The ground plane has friction $\mu_g = 0.5$. Each sequence runs for 400 steps at $\Delta t = 0.005$ with 2 substeps per step.

Per-particle attributes $\bm{C}$ consist of the normalized particle radius $\hat{r}$ and normalized internal friction $\hat{\mu}$, each standardized by mean and standard deviation across sequences. Boundary particles $(\bm{X}^{\textsc{b}}, \bm{C}^{\textsc{b}})$ are 5{,}000 samples on the ground plane with upward-pointing normals as $\bm{C}^{\textsc{b}}$. The dataset contains 400 sequences, split into 360 training, 20 validation, and 20 test.
% As illustrated in Fig.~\ref{fig:main}, our model accurately reproduces sand-ground interactions, including the progressive flattening of the sand pile, while maintaining stable contact behavior. For symmetrically initialized cases, such as the crab shown in Fig.~\ref{fig:main}, the particle interactions remain symmetric throughout the simulation. In addition, sequences initialized from previously unseen objects, such as the tiger, also produce physically plausible results, indicating that the model generalizes to different initial particle states.

\paragraph{Non-Newtonian Fluids.}
A non-Newtonian fluid flows over a sloped rigid surface, simulated with MPM~\cite{stomakhin2013material} in NVIDIA Newton~\cite{newton2025}. Fluid particles are uniformly sampled within one of 10 predefined mesh geometries and released from height $z=0.2$. The slope angle is randomized as $\theta \in [0^\circ,\,6^\circ]$, with $0^\circ$ and $6^\circ$ guaranteed for each geometry and remaining angles sampled from the same range. Material parameters are randomized per sequence: Young's modulus $E \in [9,\,15]$ and damping $d \in [60,\,90]$. Slope friction is fixed at $0.5$. Each sequence runs for 250 frames at $\Delta t = 0.01$\,s.

Per-particle attributes $\bm{C}$ consist of normalized Young's modulus $\hat{E} = 0.2(E-7)-1$ and damping $\hat{d} = 0.04(d-50)-1$, replicated across all particles. Boundary particles $(\bm{X}^{\textsc{b}}, \bm{C}^{\textsc{b}})$ are 5{,}000 samples on the sloped ground with surface normals as $\bm{C}^{\textsc{b}}$. The dataset contains 400 sequences, split into 360 training, 20 validation, and 20 test.
% As illustrated in Fig.~\ref{fig:main}, our model accurately reproduces the collision between the non-Newtonian fluid and the sloped ground, as well as the gradual post-impact sagging and deformation. The results demonstrate that the model maintains stable generalization across various simulation parameters and boundary conditions.

% \begin{figure*}
%  \centering
%  \includegraphics[width=\textwidth]{figs/non.pdf}
%  \caption{Non Hamiltonian}
%  \label{fig:non_hamiltonian}
% \end{figure*}

\paragraph{Elastic Solids.}
Elastic objects interact with a rigid sloped ground ($10^\circ$), simulated with VBD~\cite{vbd} in NVIDIA Newton~\cite{newton2025}. Each sequence uses one of 10 tetrahedralized assets with randomized initial rotation and zero initial velocity. Material parameters are fixed at Young's modulus $E=6.5\times10^5$ and Poisson's ratio $\nu=0.3$. Trajectories are 160 frames at $\Delta t=0.005$\,s with 80 substeps per frame and 150 VBD iterations per substep.

Each tetrahedral vertex is a simulated particle. Per-particle attributes $\bm{C}$ consist of normalized $(\hat{E}, \hat{\nu})$ replicated across all vertices. The topology $\mathcal{T}$ is the tetrahedral mesh connectivity. Boundary particles $(\bm{X}^{\textsc{b}}, \bm{C}^{\textsc{b}})$ are 5{,}000 samples on the sloped ground with surface normals as $\bm{C}^{\textsc{b}}$. The dataset contains 400 sequences, split into 360 training, 20 validation, and 20 test.
% Each sequence captures the interaction between a tetrahedralized elastic object and a rigid sloped ground, including contact, compression, and rebound dynamics. As illustrated in Fig.~\ref{fig:main}, our model accurately reproduces these behaviors and demonstrates consistent performance under variations in initial conditions, demonstrated by randomized object orientations, and boundary conditions, defined by varying ground slopes.

\paragraph{Proteins.}
Molecular dynamics of proteins requires extremely small timesteps (e.g., 0.5\,fs) due to stiff interatomic potentials, while biologically meaningful events such as side-chain rotations and folding occur on microsecond to millisecond timescales. Our model operates as a one-step predictor at a fixed coarser $\Delta t$, enabling rapid exploration of conformational changes without prohibitively long conventional simulations.

We generate trajectories for proteins \texttt{1a62\_A} (${\sim}$2k atoms) and \texttt{16pk\_A} (${\sim}$6k atoms) from the ATLAS dataset~\cite{ATLAS} using OpenMM~\cite{eastman2023openmm8moleculardynamics}. For each run, we first equilibrate in implicit solvent (AMBER14 + OBC1) using a Langevin integrator at $300$\,K, friction $\gamma=1.0\,\mathrm{ps}^{-1}$, and timestep $0.5$\,fs for 50{,}000 steps. From the equilibrated candidates, we select the initial state whose total energy is closest to the median among states satisfying $|E_k - \bar{E}|/|\bar{E}| \le 0.01$, where $\bar{E}$ is the mean energy over all candidates sampled every 10 steps.

Starting from the selected state, we run Hamiltonian dynamics in vacuum with the \texttt{amber14-all} force field using Velocity Verlet at $0.5$\,fs. Each rollout contains 10{,}000 integration steps, with states recorded every 100 steps, yielding trajectories of length 100. The governing equations are Hamilton's equations:
\begin{equation}
H(\bm{x},\bm{p}) = \sum_i \frac{\|\bm{p}_i\|^2}{2 m_i} + U(\bm{x}), \qquad
\dot{\bm{x}}_i = \frac{\partial H}{\partial \bm{p}_i}, \qquad
\dot{\bm{p}}_i = -\frac{\partial H}{\partial \bm{x}_i}.
\label{eq:hamilton}
\end{equation}

Per-atom attributes are $\bm{c}_i = [m_i, \tau_i]$, where $m_i$ is the atomic mass and $\tau_i \in \{0,\ldots,5\}$ is the element-type index ($\mathrm{C, H, O, N, S, P} \mapsto 0{:}5$). Topology $\mathcal{T}$ and boundary particles $(\bm{X}^{\textsc{b}}, \bm{C}^{\textsc{b}})$ are not used for this category. We repeat the procedure for 1{,}000 independent runs per protein, each with a different initial state, and split into 800 training, 100 validation, and 100 test trajectories.

% \subsection{Comparison with Baselines}
\subsection{Generalization}
\paragraph{Initial and boundary conditions.}
We use the dataset released by~\cite{deepLagrange}. Each sequence selects one of 10 box-shaped containers and initializes 1, 2, or 3 fluid bodies with randomized scales, orientations, and initial velocities, following the original data-generation protocol. Sequences are 800 frames at $\Delta t = 0.02$\,s.

Each SPH particle is a simulated particle. Per-particle attributes $\bm{C}$ consist of mass $m=0.125$ and viscosity $\nu=0.01$. Boundary particles $(\bm{X}^{\textsc{b}}, \bm{C}^{\textsc{b}})$ are sampled from the container surfaces with outward normals as $\bm{C}^{\textsc{b}}$.

\paragraph{Long rollouts.}
We simulate boundary-driven cloth twisting. Each sequence uses a triangulated square cloth at one of three resolutions ($40\!\times\!40$, $50\!\times\!50$, or $60\!\times\!60$ vertices). The boundary rotation-speed scale is randomized from $[\frac{1}{6},\frac{5}{6}]$: forces are applied to two opposite boundary strips to induce counter-rotational motion, with larger angular velocity corresponding to stronger applied force. Initial velocity is zero. Trajectories are simulated using VBD~\cite{vbd} in NVIDIA Newton~\cite{newton2025} with 4 iterations per substep, producing 250-frame sequences at $\Delta t = \frac{1}{30}$\,s with 10 substeps per frame.

Each cloth vertex is a simulated particle. Per-particle attributes $\bm{C}$ consist of per-vertex mass. The topology $\mathcal{T}$ is the triangular mesh connectivity. Boundary particles $(\bm{X}^{\textsc{b}}, \bm{C}^{\textsc{b}})$ are the two actuated boundary strips, with direction features encoding the applied twisting action. The model is trained on the first 200 frames and evaluated with auto-regressive rollout up to 250 frames.

% \begin{figure*}
%  \centering
%  \includegraphics[width=0.98\textwidth]{figs/baseline/sand_baseline.pdf}
%  \caption{Baseline comparison on the granular-sand scenario. Columns show rollout time steps (\(t=0\) to \(t=99\)); rows compare DeepLagrange, Learning to Simulate, GIOROM, our model (Ours), and ground truth. our model better preserves large-deformation collapse behavior over long horizons.}
%  \label{fig:baseline_sand}
% \end{figure*}

% \begin{figure*}
%  \centering
%  \includegraphics[width=0.98\textwidth]{figs/baseline/fluid_baseline.pdf}
%  \caption{Baseline comparison on the dam-break fluid scenario. Columns show rollout time steps (\(t=0\) to \(t=599\)); rows compare DeepLagrange, Learning to Simulate, GIOROM, our model (Ours), and ground truth. our model tracks wave propagation and obstacle interactions more faithfully over long rollouts.}
%  \label{fig:baseline_fluid}
% \end{figure*}

% \newpage

% \begin{figure*}
%  \centering
%  \includegraphics[width=0.98\textwidth]{figs/baseline/cloth_on_sphere_baseline.pdf}
%  \caption{Baseline comparison on cloth-over-sphere dynamics. Columns show rollout time steps (\(t=0\) to \(t=80\)); rows compare DeepLagrange, Learning to Simulate, MeshGraphNets, GIOROM, our model (Ours), and ground truth. our model remains stable and closest to ground-truth draping under strong contact deformation.}
%  \label{fig:baseline_cloth}
% \end{figure*}

\subsection{Ablation Studies}
\label{app:abl}
All models in this section are trained for 6 epochs, and the checkpoint with the lowest validation loss (auto-regressive rollout loss on the validation set) is selected. The default configuration uses training rollout horizon 5, spatial neighborhood radius $R_s = 0.007$, token embedding dimension 384, and encoder/decoder layers $L_e/L_d = 6/8$; its results are reported in Tab.~\ref{tab:abl_arc} for reference.

\paragraph{Training rollout horizon.}
We vary the rollout horizon in $\{2, 3, 4, 6\}$ while keeping all other settings fixed. As reported in Tab.~\ref{tab:abl_train}, longer rollouts provide stronger long-horizon supervision and reduce cumulative errors, at the cost of increased training time and memory.

\paragraph{Spatial neighborhood radius.}
The radius $R_s$ defines the support of the spatial branch in the particle tokenizer, controlling each particle's local interaction range. We vary $R_s \in \{0.002,\, 0.015,\, 0.030\}$ with other settings fixed. As shown in Tab.~\ref{tab:abl_train}, a very small radius limits neighborhood coverage and degrades local interaction modeling. Increasing $R_s$ improves rollout stability but increases the number of neighbors per particle and thus computational cost.

\input{tabs/tab_abl_train}

\paragraph{Token embedding dimension.}
We vary the embedding dimension while keeping $L_e/L_d = 6/8$ fixed. As shown in Tab.~\ref{tab:abl_arc}, increasing dimensionality initially reduces rollout error by providing greater representational capacity. Beyond a certain width, however, performance degrades, likely due to overfitting at fixed data and training budget. Wider tokens also increase computational cost.

\paragraph{Layer configuration.}
We ablate the decoder block composition and the allocation of layers between encoder and decoder. Removing self-attention or the feed-forward network from the decoder block reduces cost but increases rollout error, confirming that both components contribute to the decoder's capacity. We then vary the encoder-decoder depth split while keeping the total layer count fixed. Allocating too few layers to the encoder yields higher error due to insufficient token coarsening, while a shallow decoder limits the model's ability to transfer super-token information back to particle resolution.

\input{tabs/tab_abl_arc}

\paragraph{Interactive Control.}
Young's modulus $E \in [1.0\times10^4,\, 3.5\times10^4]$ and Poisson's ratio $\nu \in [0.24,\, 0.40]$ are randomized per sequence. External forces with magnitude $\|\bm{F}\| \in [0.7,\, 1.6]$ are applied in the positive $x$ half-plane, with direction, temporal window ($[10,\, 30]$ frames), and application region randomized. The force region is an ellipsoidal surface patch centered at an anchor sampled in normalized bounding-box coordinates $(\alpha_x, \alpha_y, \alpha_z) \in [0.18,\, 0.82] \times [0.80,\, 0.98] \times [0.18,\, 0.82]$, with patch radius in $[0.10,\, 0.18]$. Surface vertices below the anchor $y$-coordinate are pinned. Each sequence is 100 frames at $\Delta t = 1/60$\,s with 10 substeps per frame, simulated with VBD~\cite{vbd} in NVIDIA Newton~\cite{newton2025}. We use 20 meshes with 20 sequences each, totaling 400 training trajectories.

Each tetrahedral vertex except pinned surface points is a simulated particle. Per-particle attributes $\bm{C}$ consist of normalized $(\hat{E}, \hat{\nu})$ replicated across all vertices. The topology $\mathcal{T}$ is the tetrahedral mesh connectivity. Boundary particles $(\bm{X}^{\textsc{b}}, \bm{C}^{\textsc{b}})$ are the pinned surface points with outward normals as $\bm{C}^{\textsc{b}}$.

\paragraph{Inverse Design.}
We define the body particles at the final frame as $\mathcal{B} = \{ i \mid z_{T,i}(\mu) \le 0.2 \}$ and compute their mean terminal $x$-position:
\[
s(\mu) = \frac{1}{|\mathcal{B}|} \sum_{i \in \mathcal{B}} x_{T,i}(\mu).
\]
The design objective minimizes the squared distance to a target stop position $x^\star = 0.92$:
\begin{equation}
\mathcal{E}(\mu) = \bigl( s(\mu) - x^\star \bigr)^2.
\end{equation}
Gradients with respect to the friction coefficient $\mu$ are obtained by backpropagating through the full rollout. The coefficient is updated via gradient descent and mapped through a sigmoid to remain in a physically plausible range.

Training data consists of 10 sequences simulated with VBD~\cite{vbd} in NVIDIA Newton~\cite{newton2025}, varying only the friction coefficient $\mu \in [0.20,\, 0.40]$ over 10 uniformly spaced values. Other parameters are fixed: density $\rho = 1200$, Young's modulus $E = 8.0 \times 10^5$, Poisson's ratio $\nu = 0.3$, and damping $d = 0.004$. Each sequence is 99 rollout steps at $\Delta t = 0.02$\,s with 50 substeps per frame; we subsample every fifth exported frame.

Per-particle attributes $\bm{C}$ consist of the friction coefficient $\mu$, shared across all particles in a sequence. The topology $\mathcal{T}$ is the tetrahedral mesh connectivity. Boundary particles $(\bm{X}^{\textsc{b}}, \bm{C}^{\textsc{b}})$ are sampled from the ramp (10{,}000 points) and ground plane (10{,}000 points), with outward normals as $\bm{C}^{\textsc{b}}$.

%% file: tabs/tab_abl_train.tex
\begin{table}[t]
  \caption{\textbf{Hyperparameter ablation.} Effect of training rollout horizon and spatial neighborhood radius $R_s$ on the cloth task. Position and velocity errors are rollout MSE ($\times 10^{-4}$), averaged over particles, frames, and test sequences. Peak Mem.: per-GPU peak training memory. Time/Epoch: single-GPU wall-clock time.}
  \label{tab:abl_train}
  \centering
  \footnotesize
  \setlength{\tabcolsep}{4.2pt}
  \renewcommand{\arraystretch}{1.08}
  \begin{tabular*}{\linewidth}{@{\extracolsep{\fill}}lcccc@{}}
    \toprule
    Variant & Peak Mem. (GB) & Time/Ep. (h) & Pos. MSE & Vel. MSE \\
    \midrule
    \multicolumn{5}{l}{\textit{Rollout horizon}} \\
    \quad 2 & 6.89 & 0.92 & 57.43 & 94.05 \\
    \quad 3 & 11.37 & 1.74 & 5.058 & 13.18 \\
    \quad 4 & 15.86 & 2.57 & 1.816 & 4.673 \\
    \quad 6 & 20.34 & 3.36 & 0.652 & 2.424 \\
    \midrule
    \multicolumn{5}{l}{\textit{Spatial radius $R_s$}} \\
    \quad 0.002 & 20.34 & 3.29 & 0.944 & 3.289 \\
    \quad 0.015 & 20.35 & 3.52 & 0.701 & 2.018 \\
    \quad 0.030 & 20.41 & 4.10 & 0.404 & 1.062 \\
    \bottomrule
  \end{tabular*}
\end{table}

%% file: tabs/tab_abl_arc.tex
\begin{table}[t]
  \caption{\textbf{Architectural ablation.} Effect of decoder composition, token embedding dimension, and encoder/decoder layer allocation on the cloth task. Position and velocity errors are rollout MSE ($\times 10^{-4}$), averaged over particles, frames, and test sequences. Peak Mem.: per-GPU peak training memory. Time/Ep.: single-GPU wall-clock time.}
  \label{tab:abl_arc}
  \centering
  \footnotesize
  \setlength{\tabcolsep}{4.2pt}
  \renewcommand{\arraystretch}{1.08}
  \begin{tabular*}{\linewidth}{@{\extracolsep{\fill}}lcccc@{}}
    \toprule
    Variant & Peak Mem. (GB) & Time/Ep. (h) & Pos. MSE & Vel. MSE \\
    \midrule
    \multicolumn{5}{l}{\textit{Decoder composition}} \\
    \quad Full (default) & 20.33 & 3.35 & 0.831 & 3.058 \\
    \quad w/o self-attn & 12.17 & 2.39 & 10.50 & 23.48 \\
    \quad w/o FFN & 17.03 & 3.17 & 1.309 & 3.874 \\
    \quad w/o self-attn \& FFN & 8.77 & 2.24 & 14.10 & 31.44 \\
    \midrule
    \multicolumn{5}{l}{\textit{Embedding dimension}} \\
    \quad 192 & 10.19 & 1.88 & 1.006 & 3.054 \\
    \quad 256 & 14.04 & 2.17 & 0.817 & 2.664 \\
    \quad 384 (default) & 20.35 & 3.36 & 0.831 & 3.058 \\
    \quad 512 & 27.73 & 4.75 & 0.907 & 2.472 \\
    \midrule
    \multicolumn{5}{l}{\textit{Layer allocation ($L_e / L_d$)}} \\
    \quad 4 / 10 & 24.21 & 3.70 & 10.26 & 26.14 \\
    \quad 6 / 8 (default) & 20.33 & 3.35 & 0.831 & 3.058 \\
    \quad 8 / 6 & 16.41 & 3.02 & 11.23 & 27.92 \\
    \quad 10 / 4 & 18.36 & 3.19 & 20.01 & 41.33 \\
    \bottomrule
  \end{tabular*}
\end{table}